\documentclass[aps,prb,twocolumn,groupedaddress,showpacs,amsmath]{revtex4}

\usepackage{graphicx,amssymb}

\begin{document}

%

\title{Many-spin effects in inelastic neutron scattering and electron paramagnetic resonance of molecular nanomagnets}

\author{O. Waldmann}
 \email[E-mail: ]{waldmann@iac.unibe.ch}
 \affiliation{Department of Chemistry and Biochemistry, University of
Bern, CH-3012 Bern, Switzerland}

\author{H. U. G\"udel}
 \affiliation{Department of Chemistry and Biochemistry, University of Bern, CH-3012 Bern, Switzerland}

\date{\today}

\begin{abstract}
Many molecular magnetic clusters, such as single-molecule magnets, are characterized by spin ground states
with defined total spin $S$ exhibiting zero-field-splittings. In this work, the spectroscopic intensities of
the transitions within the ground-state multiplet are analyzed. In particular, the effects of a mixing with
higher-lying spin multiplets, which is produced by anisotropic interactions and is neglected in the standard
single-spin description, are investigated systematically for the two experimental techniques of inelastic
neutron scattering (INS) and electron paramagnetic resonance (EPR), with emphasis on the former technique.
The spectroscopic transition intensities are calculated analytically by constructing corresponding effective
spin operators perturbationally up to second order and consequently using irreducible tensor operator
techniques. Three main effects of spin mixing are observed. Firstly, a pronounced dependence of the INS
intensities on the momentum transfer $Q$, with a typical oscillatory behavior, emerges in first order,
signaling the many-spin nature of the wave functions in exchange-coupled clusters. Secondly, as compared to
the results of a first-order calculation, the intensities of the transitions within the spin multiplet are
affected differently by spin mixing. This allows one, thirdly, to differentiate the higher-order
contributions to the cluster magnetic anisotropy which come from the single-ion ligand-field terms and spin
mixing, respectively. The analytical results are illustrated by means of the three examples of an
antiferromagnetic heteronuclear dimer, the Mn-[3~$\times$~3] grid molecule, and the single-molecule magnet
Mn$_{12}$.
\end{abstract}


\maketitle

%

\section{Introduction}

Since the discovery of magnetic hysteresis of pure molecular origin in the molecule Mn$_{12}$ about a decade
ago,\cite{Ses93} the field of molecular nanomagnets matured into one of the most attractive research areas in
magnetism.\cite{Mn12_Fe8,tenyears,OW_CCR} In molecular nanomagnets, a defined number of metal ions are linked
by organic ligands such as to form well defined, small magnetic nanoclusters. Quite a number of spectacular
quantum effects could be observed, like quantum tunneling of the magnetization,\cite{Fri96,Tho96} quantum
phase interference effects,\cite{Wer99} quantum rotations and tunneling of the N\'eel
vector,\cite{OW_Cr8,OW_NVT} quantum magneto-oscillations,\cite{OW_QMO} or quantum superposition of high-spin
states.\cite{Bar04}

On a fundamental level, the magnetism in these clusters is described by a spin Hamiltonian, which includes
the isotropic Heisenberg exchange interactions, the ligand-field interactions, the Zeeman term, and further
anisotropic effects. In the majority of cases the situation is such, that the isotropic Heisenberg exchange
is the most dominating term, and the so-called strong-exchange limit, in which the non-Heisenberg terms are
treated in first-order perturbation theory, emerged as a standard approach to analyze their
magnetism.\cite{Ben90,Cor99,OW_FE6,OW_FE8} More precisely, a situation is considered where the dominating
Heisenberg exchange results in an energy spectrum whose interesting part consists of well separated spin
multiplets - each classified by the total-spin quantum number $S$ - which each exhibit smaller
zero-field-splittings due to the anisotropic terms in the spin Hamiltonian. A perturbational approach is
then obvious.

In the recent years, however, it became increasingly clear that the strong-exchange limit misses important
aspects and effects. For instance, it has been noted early on that the experimentally observed tunneling
splitting in single-molecule magnets can be off by orders of magnitude from that calculated from the spin
Hamiltonian describing their ground-state spin multiplets, because higher-order spin terms, most of which are
experimentally inaccessible and thus disregarded, have a crucial influence.\cite{Pro98,Wer99,Car04} The
higher-order spin terms originate from either microscopic anisotropy terms such as single-ion ligand-field
terms, or from a mixing between the different spin multiplets, which arises in second and higher order
perturbation theory (this type of mixing has been denoted e.g. as $S$ mixing in Ref.~\onlinecite{Car03}). The
origin of the magnetic anisotropy of the cluster is an obviously important question, and has been the subject
of many studies.\cite{And00,Gat01,Abb01,OW_FE8,Bir04,Sie04,Pal04,Sie05} As a result of them it turned out,
that mixing between spin multiplets can be an important source of cluster anisotropy - and for its
understanding one needs to go beyond first-order perturbation theory. Spin mixing has been observed recently
also to lead to another interesting effect. In a molecular Mn-[3~$\times$~3] grid striking quantum
magneto-oscillation were observed in the field dependence of the torque at low temperatures.\cite{OW_QMO}
These oscillations were shown to be directly related to a mixing of the levels at field-induced level
crossings,\cite{OW_QMO,Car03b} and as such may be considered as the most clear manifestation of mixing of
spin multiplets yet observed.

All these effects are intrinsically related to the fact, that the energy spectrum consists of more than one
spin multiplet, i.e., that the total spin of each spin multiplet arises as a result of the competing exchange
interactions in a many-spin system. The energy levels of the individual spin multiplets are efficiently
described by introducing effective spin Hamiltonians for each of the spin multiplets of interest, which in
first-order are naturally obtained by standard projection techniques. Here, each term in the microscopic spin
Hamiltonian corresponds to a similar term in the effective spin Hamiltonian. For instance, a single-ion
anisotropy operator such as $\hat{S}_{iz}^2$ produces a term $a(i) \hat{S}_z^2$, where $i$ indexes the spin
centers in the cluster and $a(i)$ is a projection coefficient. This approach apparently cannot grasp the
energy shifts due to the admixture of levels from the other spin multiplets. This effect leads to small
changes in the coefficients $a(i)$ and, more importantly, to new spin terms of higher-order. For example, it
is intuitively clear that a microscopic spin term $\hat{S}_{iz}^2$ adds a term proportional to
$\hat{S}_{z}^4$ to the effective spin Hamiltonian, since in second-order perturbation theory elements like
$\langle 0|\hat{S}_{iz}^2|n \rangle\langle n|\hat{S}_{iz}^2|0\rangle/\Delta$ appear, which behave like
$\hat{S}_{z}^4$ within the space of a specific spin multiplet. This mechanism has been known for long and
several methods have been developed to extend the effective spin Hamiltonian concept beyond first-order.
Recently, Liviotti {\it et al.} presented a rather appealing strategy to incorporate the spin-mixing effects
in the effective Hamiltonian up to second-order in perturbation theory.\cite{Car03}

By construction, the effective spin Hamiltonian is designed to reproduce the energy levels of a spin
multiplet correctly, and thus allows to describe thermodynamic measurements or the transition energies in
spectroscopic work with high accuracy. However, this alone is not sufficient to ensure that also the
transition matrix elements, which determine spectroscopically observed intensities, are reproduced
correctly. To see this, let us consider a specific spin multiplet. Due to spin mixing, a small amount of the
wave functions from the other spin multiplets is mixed in, and symbolically we write $|0\rangle =
|\bar{0}\rangle + \varepsilon |\bar{n}\rangle$, where $\varepsilon$ is a measure for the mixing
($|\bar{0}\rangle$ and $|\bar{n}\rangle$ denote unperturbed wave functions). Equivalently one may write
$|0\rangle = T |\bar{0}\rangle$, with the transformation $T$ chosen appropriately. Instead of calculating
the energies from $\langle0|\hat{H}|0\rangle$, it is the idea of the effective Hamiltonian concept to
calculate it from $\langle\bar{0}|T^\dagger \hat{H} T|\bar{0}\rangle =
\langle\bar{0}|\hat{\bar{H}}|\bar{0}\rangle$, with the effective Hamiltonian $\hat{\bar{H}} \equiv
T^\dagger\hat{H}T$ (a short but excellent review of the effective Hamiltonian concept is found in
Ref.~\onlinecite{Cal04}). One may express it this way: physically the wave functions are modified by the
mixing, but the effect with respect to the energies can be described within a framework, which keeps the
Hilbert space fixed but introduces a modified Hamiltonian so as to ensure $\langle0|\hat{H}|0\rangle =
\langle\bar{0}|\hat{\bar{H}}|\bar{0}\rangle$. Since the enlargement of the space which is required to
account for the mixing cannot be accommodated by the zero-order wave functions, it is intuitively not
surprising that the Hamilton operator in turn has to be adapted, or extended, respectively. However, the
very same is true for any operator $\hat{O}$ if its matrix elements are to be calculated within the
restricted space of a spin multiplet. Therefore, similar as for the Hamiltonian, the operator $\hat{O}$ has
to be substituted by an effective operator $\hat{\bar{O}}$, in which additional terms with no direct
counterparts appear, in order to ensure $\langle n|\hat{O}|m\rangle =
\langle\bar{n}|\hat{\bar{O}}|\bar{m}\rangle$. In short, if a mixing between spin multiplets is
non-negligible, the operators which determine the spectroscopic intensities need to be adapted or extended
too, in order to yield correct results. At this point it is worth mentioning that in a complete numerical
diagonalization these issues of course do not arise, since the exact eigen functions obtained this way
trivially include the mixing of different multiplets.

In this work, we will analyze the effects of spin mixing on the spectroscopic intensities for transitions
within a zero-field-split spin multiplet. This task essentially involves the construction of the
corresponding effective operators, which is achieved by a procedure similar to that used by Liviotti {\it et
al.}\cite{Car03} for the construction of the effective spin Hamiltonian (and, as a spin-off result, we also
provide a considerable improvement of their method). The discussion will be conducted with the specific
examples of the single-molecule magnets, or more generally the spin clusters with non-compensated
ground-state spin, in mind. For this class of compounds, the magnetic excitations within their ground-state
multiplets have been investigated intensely.\cite{Cac98,Mir99,And00,Amo01,Bas03} Such studies provide
detailed information on the magnetic anisotropy, which is of obvious importance for the understanding of
their quantum properties. These systems, in fact, represent canonical examples, and the previous studies on
the magnetic excitations within their high-spin ground states largely motivated this work. It will be shown,
that a careful analysis of the transition intensities can yield crucial information not obtainable
otherwise. As a simple example, from the energy spectrum alone it is not possible to decide whether a term
$\hat{S}_{z}^4$ in the effective spin Hamiltonian arises as a second-order effect from the single-ion terms
$\hat{S}_{iz}^2$, or as a first-order result from single-ion terms $\hat{S}_{iz}^4$. The spectroscopic
intensity in principle allows to discriminate these two origins.

At low temperatures, where only the ground-state multiplet is relevant, these systems behave like a single
spin. The corresponding effective spin Hamiltonian is known as single-spin model or giant-spin model. The
transitions within the spin ground states were investigated experimentally by inelastic neutron scattering
(INS) and electron paramagnetic resonance (EPR). In both techniques, the measured intensities are related to
the transition matrix elements of the local spin operators, $\langle n|\hat{S}_{i\alpha}|m \rangle$ ($\alpha
= x,y,z$). In analogy to the spin density map, which corresponds to $\langle n|\hat{S}_{i\alpha}|n \rangle$,
we will call the set of matrix elements $\langle n|\hat{S}_{i\alpha}|m \rangle$ the spin transition map.
Clearly, the topology of the spin transition map is different for each transition. The effect of the spin
mixing on the ground-state intensities may then be understood intuitively as the admixing of spin transition
maps with topologies different from that of the ground state.

The discussion will focus mainly on the situation encountered in INS, as it is, when affordable, the
technique of choice since the analysis is not affected by the complexities introduced by an applied magnetic
field; the discussion for EPR will be limited to some important general issues.\cite{FDMRS} INS is also of
special interest since it allows to probe the local transition matrix elements via the dependence of the
scattering intensity on momentum transfer ($Q$ dependence), i.e., one may say that INS allows to probe the
topology of the spin transition map, while EPR relates to an average. In spin clusters, the $Q$ dependence
exhibits typical oscillations due to so-called interference factors,\cite{Fur77,OW_INS} which are absent for
mononuclear compounds and are thus intrinsically a many-spin effect. However, the INS transitions within the
ground-state multiplets of high-spin clusters have been analyzed so far as if these systems were true single
spins, neglecting the $Q$ dependence (the only exception we are aware of is Ref.~\onlinecite{Hen97}). This
approach, denoted as single-spin approach henceforth, leads to some subtleties, which will be addressed.

The manuscript is organized as follows. First, in the next section, the single-spin approach is surveyed as
this sets the context for the remainder of the work. Then, in section~\ref{sGeneral}, the problem is
carefully stated and the notation set up. Also, some important relations are given, in particular the very
useful eq.~(\ref{Vexpand}), which is key to this work. The formalism developed for the calculation of
spin-mixing effects on the transition intensities turned out to be very efficient also for the construction
of the effective spin Hamiltonian, which is considered in section~\ref{sHeff}, where also a general result is
derived. In section~\ref{sINS} then, which represents the core of this work, the effects of spin-mixing on
the INS transitions are discussed theoretically and by means of some examples. Section~\ref{sEPR} is devoted
to a discussion of some effects of mixing in EPR. Section~\ref{sConclusions} finally presents a conclusion.
Some supporting information, helpful for the understanding of the formalism and for actual calculations, is
provided in the appendix. Readers who are mainly interested in a qualitative understanding of the effects
may skip the technical subsections marked by an asterisk.

\section{Survey of the single-spin approach for analyzing INS data of high-spin clusters}
\label{sSSApproach}

The INS transitions within the ground-state spin multiplet of high-spin clusters were analyzed so far using a
formula derived for mononuclear compounds (with the exception of Ref.~\onlinecite{Hen97}). For a mononuclear
paramagnet, the transition intensity from state $|n\rangle$ to $|m\rangle$, $I_{nm}$, is proportional to the
matrix element $|\langle n |\hat{S}_{\perp}| m\rangle|^2$, where $\hat{S}_{\perp}$ is the spin component
perpendicular to the scattering vector ${\bf Q} = {\bf k} - {\bf k}'$.\cite{Bir72} For a powder sample,
after averaging over all orientations of ${\bf Q}$, one obtains
\begin{equation}
\label{Iss}
  I_{nm} \propto p_n \left( 2 |\langle n |\hat{S}_z| m\rangle|^2  + |\langle n |\hat{S}_{+}| m\rangle|^2
  + |\langle n |\hat{S}_{-}| m\rangle|^2 \right)
\end{equation}
($p_n$ is the Boltzmann factor for the state $n$).\cite{Cac98,Mir99}

The use of this formula was certainly motivated by the insight, that high-spin clusters in many respects
behave at low temperatures like large, single spins. However, using eq.~(\ref{Iss}), which is the
single-spin approach, can produce substantially erroneous results for the calculated INS intensities.

It is a celebrated feature of the INS intensity to provide information about the wave functions (for a recent
discussion see Ref.~\onlinecite{OW_INS} and the references cited therein). It is thus intuitively clear, that
the INS intensity "feels" the differences between a single-spin and a many-spin wave function. In fact, for a
polynuclear spin cluster the INS intensity depends on the local transition matrix elements and the
geometrical arrangement of the spin centers in the cluster via a factor $\exp[i{\bf Q}({\bf R}_i - {\bf
R}_j)]\langle n |\hat{S}_{i\alpha}| m\rangle\langle m |\hat{S}_{j\beta}|n\rangle$ (${\bf R}_i$ denotes the
position vector of the $i$th spin center). This gives rise to characteristic interference factors in the INS
cross section of powder samples, and hence characteristic $Q$ dependencies of the INS intensity, which are
typically oscillatory.\cite{Fur77,OW_INS} For a single spin, in contrast, such interference effects are
absent, and the INS intensity is independent of $Q$, see eq.~(\ref{Iss}). Thus, as mentioned already in the
introduction, the interference effects visible in the $Q$ dependencies are a direct fingerprint of the
many-spin nature of the wave functions in polynuclear clusters. These effects were indeed observed for, e.g.,
the single-molecule magnet Mn$_{12}$,\cite{Hen97} showing that the deficit of the single-spin approach to
reproduce the $Q$ dependence is experimentally relevant.

In a first approximation, the $Q$ dependence is the same for all transitions within a spin multiplet. In
fact, in section~\ref{sINS} we will show that this is strictly correct up to first-order in perturbation
theory. Thus, if one is analyzing the transition intensities at constant $Q$, the relative intensities of the
transitions as calculated from eq.~(\ref{Iss}) are reasonably accurate. However, due to obvious advantages,
the spectra are often recorded on time-of-flight (TOF) spectrometers, and the integrated spectra, which are
obtained by summing the data of all detector banks, are used in the analysis. In that case a rather trivial
but important problem occurs, which we mention here since it seems not to have been considered in the
previous analyses:

In the integrated TOF spectra, the intensity of each peak corresponds to $I_{nm}^{tot} =
\int_{Q_{min}}^{Q_{max}} g(Q) I_{nm}(Q) dQ$.\cite{TOF} Thus, even if the INS intensities are independent on
$Q$, the TOF intensities are not simply proportional to the squared matrix elements and the Boltzmann factor,
but also scale with $G(E) \equiv \int_{Q_{min}}^{Q_{max}} g(Q) dQ$. Since $G(E)$, however, depends only on
instrumental parameters,\cite{TOF} it is easy enough to account for that (though this has been disregarded in
previous works using the single-spin approach). But this procedure is not working if the $Q$ dependence
exhibits pronounced maxima and minima, as it is often the case. Then, for one peak $Q_{min}$ and $Q_{max}$
may be such as to bracket just the top of a maximum, while for another peak they may bracket also the nearby
minima - with corresponding effects on the integrated peak intensities. In conclusion, when using the
single-spin approach for the analysis of integrated TOF spectra one should be aware of a possibility of
significant deviations between experimental and calculated intensities. As a side-effect, the TOF intensities
of the Stokes and anti-Stokes peaks will not obey the relationship $I(-\omega) = \exp(\hbar\omega/k_BT)
I(\omega)$. The situation can be improved considerably if a rough approximation for the $Q$ dependence is
available, e.g., as extracted from the TOF data itself, or if approximations for the wave functions are
conceivable. In the remainder of this work, we will disregard these issues, but focus entirely on the effects
of the many-spin nature of the wave functions on the intensities.

Finally it is mentioned that the above considerations are rather specific to INS. For instance, the EPR
signal does not provide information about the spatial properties of the involved wave functions as it uses
spatially homogeneous magnetic fields.\cite{EPR1} Accordingly, the single-spin model works much better for
the EPR intensity. In the language of perturbation theory, the $Q$ dependence of the INS intensity is a
first-order effect of the many-spin nature of the wave functions, while the EPR intensity is affected only in
second-order ({\it vide infra}).

%

\section{General relations}
\label{sGeneral}

The microscopic spin Hamiltonian describing a magnetic spin cluster of $N$ spin centers, where the $i$th
center has spin length $S_i$, is
\begin{eqnarray}
\label{H}
 \hat{H} &=& - \sum_{i\neq j} J_{ij} \hat{{\bf S}}_i \cdot \hat{{\bf S}}_j
       + \sum_{i\neq j} \hat{{\bf S}}_i \cdot {\bf \Pi}_{ij} \cdot \hat{{\bf S}}_j \cr &&
       + \sum_{i}  \hat{{\bf S}}_i \cdot {\bf D}_{i} \cdot \hat{{\bf S}}_i
       + \sum_{i} \sum_{m} B^m_4(i) \hat{O}^m_4(i) \cr &&
       + \mu_B \sum_{i} \hat{{\bf S}}_i \cdot {\bf g}_{i} \cdot {\bf B}.
\end{eqnarray}
The first term represents the isotropic exchange interactions, the second term describes anisotropic and
antisymmetric exchange and dipole-dipole interactions, the third and fourth terms are single-ion ligand
field terms, and the last one is the Zeeman interaction. This expression is quite general; from the usually
considered terms only the higher-order exchange terms like biquadratic exchange were excluded (and this only
for convenience). The tensors ${\bf D}_{i}$ are traceless and symmetric, $\hat{O}^m_4(i)$ are Stevens
operators acting on the $i$th spin,\cite{Abr70} and $\mu_B$ is the Bohr magneton.

In this work we are considering systems in which the isotropic exchange interaction dominates the other
terms. Then the basis states of the Hilbert space are conveniently chosen as the spin states $|\kappa S M
\rangle$, where $\kappa$ abbreviates the intermediate spin values according to the chosen coupling scheme.
$S$ and $M$ denote the total spin and magnetic quantum number, respectively. Working in this basis, the
irreducible tensor operator (ITO) technique provides a powerful tool for the calculation of matrix
elements.\cite{Ben90,ITO} In terms of ITOs, the above Hamiltonian writes
\begin{eqnarray}
\label{Hito}
 \hat{H} &=& -\sum_{i\neq j} J_{ij} \hat{{\bf S}}_i \cdot \hat{{\bf S}}_j
       + \sum_{i\neq j,q} T^{(2)*}_{q}({\bf \Pi}_{ij}) \hat{T}^{(2)}_q(ij) \cr &&
       + \sum_{i,q} T^{(2)*}_{q}({\bf D}_{i}) \hat{T}^{(2)}_q(i)
       + \sum_{i,q} T^{(4)*}_{q}(B_4(i)) \hat{T}^{(4)}_q(i) \cr &&
       + \sum_{i,q} T^{(1)*}_{q}({\bf b}_i) \hat{T}^{(1)}_q(i).
\end{eqnarray}
Here, ${\bf b}_i = \mu_B {\bf g}_i \cdot {\bf B}$ was introduced. $T^{(k)}_q({\bf V})$ denotes the $q$th
component of the irreducible spherical tensor of rank $k$ related to the Cartesian tensor ${\bf V}$ [it is
proportional to the spherical harmonics $Y_{kq}({\bf V})$, and obeys $T^{(k)*}_q = (-1)^q T^{(k)}_{-q}$ ].
$\hat{T}^{(k)}_q(i)$ denotes an irreducible tensor operator which acts in the spin space of the spin center
$i$, and $\hat{T}^{(k)}_q(ij)$ denotes an ITO which is composed of two single-spin ITOs acting in the spin
spaces of the spins $i$ and $j$ with $i \neq j$, respectively.\cite{ITO2} In equation~(\ref{Hito}), each term
of eq.~(\ref{H}) is replaced by a corresponding spherical scalar product, expressing the fact that a
Hamiltonian transforms under rotations in space as a scalar (see appendix~\ref{app_ITO} for more details
concerning the Cartesian and spherical representations of a spin Hamiltonian).

The second and third terms in eq.~(\ref{H}) or eq.~(\ref{Hito}), respectively, can be combined into
$\sum_{ij} \hat{{\bf S}}_i \cdot {\bf \Lambda}_{ij} \cdot \hat{{\bf S}}_j = \sum_{ijq} T^{(2)*}_{q}({\bf
\Lambda}_{ij}) \hat{T}^{(2)}_q(ij)$ if one defines ${\bf \Lambda}_{ij} = {\bf \Pi}_{ij} (1-\delta_{ij}) +
{\bf D}_i \delta_{ij}$, and adopts the convention that $\hat{T}^{(k)}_q(ii)$ means $\hat{T}^{(k)}_q(i)$. This
will simplify calculations considerably as the bilinear anisotropy terms in $\hat{H}$ can be treated
simultaneously this way.

Many results in this work depend only on the transformation properties of the ITOs, and not on $ij$ or $i$,
i.e., on which spin centers are involved. We often simplify notation by writing $\hat{T}^{(k)}_q(s)$, where
$s$ stands for either $ij$ or $i$ (and similarly in sums like $\sum_s$). If the meaning is obvious from the
context, we also write just $\hat{T}^{(k)}_q$.

The perturbative approach which is adopted to calculate energies and transition matrix elements is as
follows. The unperturbed Hamiltonian consists of the isotropic exchange interactions, $\hat{H}_0 = -
\sum_{ij} J_{ij} \hat{{\bf S}}_i \cdot \hat{{\bf S}}_j$. The corresponding zero-order eigenstates are
written as $|\tau S M\rangle$, where $\tau$ denotes additional quantum numbers.

The aim is now the following: Within one of the (non-degenerate) spin multiplets specified by $\tau S$, the
microscopic spin Hamiltonian $\hat{H}$ shall be replaced by an effective spin Hamiltonian operating in a
single-spin space $|S M\rangle$. This is achieved in two steps. First one determines the effective
Hamiltonian $\hat{\bar{H}}$ acting in the subspace $|\tau S M\rangle$ by a perturbation technique, and then
one expresses $\hat{\bar{H}}$ by single-spin operators using ITO techniques. Similarly, the spin operators
$\hat{S}_{i\alpha}$ shall be replaced by effective spin operators $\hat{\bar{S}}_{i\alpha}$ expressed by
single-spin operators $\hat{S}_\alpha$. In the following subsections we provide some background as needed in
this work, and present the very useful eq.~(\ref{Vexpand}) (readers not interested in details may skip these
technical sections).

\subsection{Effective Hamiltonian, effective operators*}
\label{sGeneralA}

In the spectral representation, the Hamilton operator $\hat{H}$ becomes
\begin{equation}
 \hat{H} = \sum{ |n\rangle \langle n|\hat{H}|n\rangle \langle n| },
\end{equation}
where $|n\rangle$ are the eigenfunctions of $\hat{H}$. The effective Hamiltonian $\hat{\bar{H}}$ describing a
subset of levels of $\hat{H}$ is then constructed by introducing a model space $\mathbb{S}$, the states of
which will be denoted as $|\bar{n}\rangle$, with a one-to-one correspondence to the subset of levels of
$\hat{H}$, i.e., $n \leftrightarrow \bar{n}$. Then
\begin{equation}
\hat{\bar{H}} = P_\mathbb{S} \hat{H} P_\mathbb{S} =  \sum{ |\bar{n}\rangle \langle n|\hat{H}|n\rangle \langle
\bar{n}| },
\end{equation}
where $P_\mathbb{S} = \sum{ |\bar{n}\rangle \langle n|}$ was introduced. Frequently, as is the case in this
work, the model space $\mathbb{S}$ is a subspace of the Hilbert space of $\hat{H}$, then the procedure
consists of finding an unitary transformation with $|\bar{n}\rangle = T |n\rangle$:
\begin{equation}
 \langle n|\hat{H}|n\rangle =  \langle \bar{n}| T^\dag \hat{H} T |\bar{n}\rangle = \langle \bar{n}| \hat{\bar{H}}
 |\bar{n}\rangle,
\end{equation}

Several methods are available to find the transformation $T$
perturbatively.\cite{Pry50,Blo58,Clo60,Sch66,Tak77,Sch90} As usual, one starts from $\hat{H} = \hat{H}_0 +
\hat{H}_1$ with $\hat{H}_1$ being small. In general, the spectrum of $\hat{H}_0$ exhibits degeneracies. The
eigenstates of $\hat{H}_0$ are thus written as $|l \mu\rangle$ such that $\hat{H}_0 |l \mu\rangle = E_l |l
\mu\rangle$. The different methods lead to equivalent results in the order considered here (the results below
seem to have been first reported by Pryce \cite{Pry50}). Following Schrieffer and Wolff,\cite{Sch66} or
Schlichter,\cite{Sch90} one chooses $T = e^{iU}$, which implies $\hat{\bar{H}} \cong \hat{H} + i[\hat{H},U] -
1/2 [ [\hat{H},U],U]$, and separates the perturbation as $\hat{H}_1 = \hat{H}_a + \hat{H}_b$ such that
$\langle l \mu|\hat{H}_a|l' \mu'\rangle = 0$ for $l \neq l'$ and $\langle l \mu|\hat{H}_b|l' \mu'\rangle = 0$
for $l = l'$. Then, the matrix elements of $\hat{H}_b$ can be minimized by choosing $\hat{H}_b +  i [
\hat{H}_0+\hat{H}_a, U ] = 0$, which results in
\begin{equation}
 \hat{\bar{H}} =  \hat{H}_0 + \hat{H}_a + i/2 [ \hat{H}_b, U ]
\end{equation}
up to second order in $\hat{H}_b$ and/or $U$, respectively. The transformation $U$ is obtained as
\begin{equation}
 U =  {1\over i} \sum_{l \neq l'} { P_l \hat{H}_b P_{l'} \over E_{l'}-E_l },
\end{equation}
and the effective Hamiltonian acting in the subspace $P_l$ as
\begin{equation}
\label{Heff}
 \hat{\bar{H}} = P_l \left( \hat{H}  - \sum_{l' \neq l}{ \hat{H} P_{l'} \hat{H} \over E_{l'}-E_l } \right) P_l,
\end{equation}
with $P_l = \sum_\mu |l \mu\rangle \langle l \mu|$. Note that $P_l \hat{H} P_{l'} = P_l \hat{H}_b P_{l'}$
for $l \neq l'$.

Similarly as for the Hamiltonian, an arbitrary operator $\hat{O}$ has to be replaced too by an effective
operator $\hat{\bar{O}}$ acting in the model space $\mathbb{S}$ in order to ensure $\langle
n|\hat{O}|m\rangle = \langle \bar{n}| \hat{\bar{O}}|\bar{m}\rangle$. Accordingly,
\begin{equation}
 \hat{\bar{O}} =  P_\mathbb{S} \hat{O} P_\mathbb{S} = \sum{ |\bar{n}\rangle \langle n|\hat{O}|m\rangle \langle \bar{m}| },
\end{equation}
or, if the model space $\mathbb{S}$ is again a subspace of the Hilbert space of $\hat{H}$,
\begin{equation}
 \hat{\bar{O}} = e^{-iU} \hat{O} e^{iU}.
\end{equation}
In contrast to the result for the effective Hamiltonian, the effective operator is obtained as $\hat{\bar{O}}
= \hat{O} + i [ \hat{O}, U ]$. The difference occurs because $\hat{H}_b$ is of lower order than $\hat{O}$.
Inserting the above perturbation result for the transformation $U$, one finally obtains
\begin{equation}
\label{Oeff}
 \hat{\bar{O}} = P_l \left( \hat{O}  - \sum_{l' \neq l}{ \hat{H} P_{l'} \hat{O} +  \hat{O} P_{l'} \hat{H} \over E_{l'}-E_l } \right) P_l.
\end{equation}

In the context of this work, having written the zero-order eigenstates as $|\tau S M\rangle$, the index $l$
corresponds to $\tau S$, and $\mu$ to $M$. The operators obtained from eqs.~(\ref{Heff}) or (\ref{Oeff}),
which are acting in the subspace $|\tau S M\rangle$ with $\tau$ and $S$ fixed, can be replaced, as will be
discussed in the next subsections, by operators acting in the single-spin space $|S M\rangle$.

In order to avoid confusion which spin space is meant, we apply the following conventions: Whenever both
quantum numbers $\tau$ and $S$ appear in a matrix or reduced matrix element, it should be understood as a
many-spin matrix element, while if only $S$ appears, a single-spin matrix element is meant. If not already
unequivocally clarified by this, $\hat{T}^{(k)}_q(s)$ indicates a many-spin ITO acting in the space $|\tau S
M\rangle$, and $\hat{T}^{(k)}_q(S)$ a single-spin ITO acting in the space $|S M\rangle$.

\subsection{Expansion of a spin operator into ITOs*}
\label{sGeneralB}

The aim is to find the expansion of an operator $\hat{O}(S)$, which is a function of spin operators, into
irreducible tensor operators:
\begin{equation}
\label{o2ito}
 \hat{O}(S) =  \sum_{kq} c_{kq} \hat{T}^{(k)}_q(S).
\end{equation}
The coefficients can be calculated by considering the matrix elements $\langle S M | \hat{O} | S M' \rangle$
within a spin multiplet. With the help of the Wigner-Eckardt theorem and the orthogonality relation
\begin{eqnarray}
\nonumber
 \sum_{m_1 m_2} \left(\begin{array}{ccc} j_1&j_2&j_3 \\ m_1&m_2&m_3 \end{array} \right)
 \left(\begin{array}{ccc} j_1&j_2&j'_3 \\ m_1&m_2&m'_3 \end{array} \right)
 = \cr
 {1 \over 2 j_3 +1 } \delta_{j_3 j'_3} \delta_{m_3 m'_3},
\end{eqnarray}
where $(...)$ denotes a Wigner-3$j$ symbol, one directly obtains the following useful result:
\begin{eqnarray}
\label{ckq}
 c_{kq} = { 2 k + 1 \over \langle S \| \hat{T}^{(k)} \| S \rangle }
 \cr\cr \times
 \sum_{M M'}(-1)^{M-S} \left(\begin{array}{ccc} S&k&S \\ -M&q&M'  \end{array} \right) \langle S M | \hat{O} | S M'
 \rangle.
\end{eqnarray}
Here, $\langle S \| \hat{T}^{(k)} \| S \rangle$ denotes the (single-spin) reduced matrix element. If the
operator $\hat{O}$ is acting in a many-spin space, $\hat{O}(S)$ should be interpreted as $\sum_{MM'} | S M
\rangle \langle \tau S M | \hat{O} | \tau S M' \rangle \langle S M'|$.

\subsection{Spin operator equivalents*}
\label{sGeneralC}

The procedure to find the spin operator equivalents of the first-order terms $P_{\tau S} \hat{O} P_{\tau S}$
is well known,\cite{Ben90} and is reproduced here for convenience. One first expresses $\hat{O}$ by
(many-spin) ITOs, similar as in eq.~(\ref{o2ito}). Then the Wigner-Eckardt theorem is exploited resulting in
\begin{eqnarray}
\label{Texpand}
 P_{\tau S} \hat{T}^{(k)}_{q}(s) P_{\tau S} &=& \Gamma_{k}(s) \: \hat{T}^{(k)}_q(S),
 \cr\cr
 \Gamma_{k}(s) &=& { \langle \tau S \| \hat{T}^{(k)}(s) \| \tau S \rangle \over \langle S \| \hat{T}^{(k)} \| S \rangle },
\end{eqnarray}
with the projection coefficients $\Gamma_{k}(s)$ (in case of ambiguities, $\Gamma_{k}$ will be called a
$1^{st}$-order projection coefficient).

To find the spin operator equivalents for expressions of the form $P_{\tau S} \hat{O}_1 P_{\tau' S'}
\hat{O}_2 P_{\tau S}$, which appear in second order, one again first decomposes the operators $\hat{O}_1$
and $\hat{O}_2$ into ITOs. This leaves one with the evaluation of terms alike $P_{\tau S}
\hat{T}^{(k_1)}_{q_1} P_{\tau' S'} \hat{T}^{(k_2)}_{q_2} P_{\tau S}$. Though they could be replaced directly
by effective single-spin operators (which, however, are not easily found), it is more convenient to expand
$P_{\tau S} \hat{T}^{(k_1)}_{q_1} P_{\tau' S'} \hat{T}^{(k_2)}_{q_2} P_{\tau S}$ in turn into single-spin
ITOs because, as will be shown below, the respective expansion coefficients can be calculated very easily.

The matrix elements of $P_{\tau S} \hat{T}^{(k_1)}_{q_1} P_{\tau' S'} \hat{T}^{(k_2)}_{q_2} P_{\tau S}$
within the spin multiplet labelled by $\tau S$ are
\begin{eqnarray}
\nonumber
 \sum_{M'} \langle \tau S M | \hat{T}^{(k_1)}_{q_1} | \tau' S' M' \rangle \langle \tau' S'
  M' |  \hat{T}^{(k_2)}_{q_2} | \tau S M'' \rangle =
 \cr\cr
 \langle \tau S \| \hat{T}^{(k_1)} \| \tau' S' \rangle \langle \tau' S' \| \hat{T}^{(k_2)} \| \tau S \rangle
 \sum_{M'} (-1)^{S-M+S'-M'}
 \times \cr\cr
 \left(\begin{array}{ccc} S&k_1&S' \\ -M&q_1&M' \end{array} \right)
 \left(\begin{array}{ccc} S'&k_2&S \\ -M'&q_2&M'' \end{array} \right).
\end{eqnarray}
Insertion of this expression into eq.~(\ref{ckq}) in order to get the expansion coefficients requires the
evaluation of

\begin{eqnarray}
\label{sum3m}
 \sum_{M M' M''} (-1)^{S'-M'} \left(\begin{array}{ccc} S&k&S \\ -M&q&M'' \end{array} \right)
 \cr\cr \times
 \left(\begin{array}{ccc} S&k_1&S' \\ -M&q_1&M' \end{array} \right)
 \left(\begin{array}{ccc} S'&k_2&S \\ -M'&q_2&M'' \end{array} \right),
\end{eqnarray}
which, by using some symmetry properties of the Wigner-3$j$ symbols and $(-1)^{-M'} =
(-1)^{M'+M+M''-q_1+q_2}$, can be transformed into
\begin{eqnarray}
\nonumber
 (-1)^{S'-q_1+q_2} \sum_{M M' M''} (-1)^{M'+M+M''}
 \left(\begin{array}{ccc} S'&S&k_1 \\ M'&-M&q_1 \end{array} \right)
 \cr\cr \times
 \left(\begin{array}{ccc} S&S&k \\ M&-M''&-q \end{array} \right)
 \left(\begin{array}{ccc} S&S'&k_2 \\ M''&-M'&q_2  \end{array} \right).
\end{eqnarray}
With the help of a relation between the Wigner-3$j$ and 6$j$ symbols,
\begin{eqnarray}
\nonumber
 \sum_{M_1 M_2 M_3} (-1)^{J_1+J_2+J_3+M_1+M_2+M_3}
 \left(\begin{array}{ccc} J_1&J_2&j_3 \\ M_1&-M_2&m_3 \end{array} \right)
 \cr\cr \times
 \left(\begin{array}{ccc} J_2&J_3&j_1 \\ M_2&-M_3&m_1 \end{array} \right)
 \left(\begin{array}{ccc} J_3&J_1&j_2 \\ M_3&-M_1&m_2 \end{array} \right) =
 \cr\cr
 \left(\begin{array}{ccc} j_1&j_2&j_3 \\ m_1&m_2&m_3 \end{array} \right)
 \left\{\begin{array}{ccc} j_1&j_2&j_3 \\ J_1&J_2&J_3 \end{array} \right\},
\end{eqnarray}
where $\{...\}$ denotes a Wigner-6$j$ symbol, eq.~(\ref{sum3m}) assumes the simple expression
\begin{eqnarray}
\nonumber
  (-1)^{-2S -q_1+q_2} \left( \begin{array}{ccc} k&k_2&k_1 \\ -q&q_2&q_1 \end{array} \right)
  \left\{ \begin{array}{ccc} k&k_2&k_1 \\ S'&S&S \end{array} \right\}.
\end{eqnarray}
Using $\langle \tau S \| \hat{T}^{k} \| \tau' S' \rangle = (-1)^{S'-S} \langle \tau' S' \| \hat{T}^{(k)\dag}
\| \tau S \rangle^*$ and introducing the coefficients
\begin{eqnarray}
\label{d}
 c_{k_1 k_2 k} &=&
 (-1)^{S+S'} (2 k + 1) \langle S \| \hat{T}^{(k)} \| S \rangle
 \left\{ \begin{array}{ccc} k_1&k_2&k \\ S&S&S' \end{array} \right\},
 \cr\cr
 d^{k_1 k_2 k}_{q_1 q_2  q} &=&
 c_{k_1 k_2 k} (-1)^{q} \left(\begin{array}{ccc} k_1&k_2&k \\ -q_1&-q_2&q \end{array} \right),
 \cr\cr
 \Gamma_{k_1k}(s) &=& { \langle \tau S \| \hat{T}^{(k_1)}(s) \| \tau' S' \rangle
  \over
 \langle S \| \hat{T}^{(k)} \| S \rangle },
\end{eqnarray}
where we call $\Gamma_{k_1k}(s)$ a $2^{nd}$-order projection coefficient, a very useful formula is finally
obtained:
\begin{eqnarray}
\label{Vexpand}
 P_{\tau S} \hat{T}^{(k_1)}_{q_1}(s_1) P_{\tau' S'} \hat{T}^{(k_2)}_{q_2}(s_2) P_{\tau S} =
 \cr\cr
 \sum_{k} \Gamma_{k_1 k}(s_1)  \Gamma^*_{k_2 k}(s_2) d^{k_1 k_2 k}_{q_1 q_2  q} \hat{T}^{(k)}_q(S).
\end{eqnarray}

Here, the sum runs only over $k$ because of the restriction $q= q_1+q_2$. The similar structure of
eq.~(\ref{Vexpand}) and eq.~(\ref{Texpand}) is satisfying.

The coefficients $d^{k_1 k_2 k}_{q_1 q_2 q}$ do not depend on the actual wave functions involved and thus can
be calculated, and tabulated, once and for all. The potentially most needed coefficients are listed in
Table~\ref{tab1}. For calculations it is useful to realize that with respect to the first two "columns" they
have the same symmetry properties as the Wigner-3$j$ symbols, e.g., $d^{k_2 k_1 k}_{q_2 q_1 q} =
(-1)^{k_1+k_2+k} d^{k_1 k_2 k}_{q_1 q_2 q}$. It is also useful to note that for a given $k_1$ the
$2^{nd}$-order projection coefficients for different $k$ are trivially related to each other as they all
refer to the same reduced matrix element, e.g., $\Gamma_{1,k}(s) = \langle S \| \hat{T}^{(1)} \| S \rangle /
\langle S \| \hat{T}^{(k)} \| S \rangle \Gamma_{1,1}(s)$.

\begin{table*}
\caption{\label{tab1}The coefficients $d^{k_1k_2k}_{000}$ discussed in the text for the potentially most need
cases. Independent on $S'-S$ is $d^{220}_{000} = 1/5$. }
\begin{ruledtabular}
\begin{tabular}{lcccc}
$S'$ & $d^{222}_{000}$ & $d^{224}_{000}$ & $d^{121}_{000}$ & $d^{123}_{000}$ \\
\hline
$S-2$  & $-{\sqrt{6} \over 21 } (S+1)(2S+3)$  &  ${3 \sqrt{70} \over 2450}  (S+1)(2S+3)(S+2)(2S+5)$ &-&-\\
$S-1$  & ${\sqrt{6} \over 42}   (S-5)(2S+3)$    &  $-{ 3 \sqrt{70} \over 1225} (2S-3)(2S+3)(S+2)(2S+5)$ &
${\sqrt{3} \over 5} \sqrt{(S-1)(S+1)}$  & $- d^{121}_{000} {1 \over \sqrt{10}} (2S+3)(S+2)$ \\
$S$    & ${\sqrt{6} \over 42}   (2S-3)(2S+5)$   &  ${9 \sqrt{70} \over 1225} (2S-3)(S-1)(S+2)(2S+5)$ & ${1
\over 5} \sqrt{(2S-1)(2S+3)}$  & $ d^{121}_{000} {3 \over \sqrt{10}} (S-1)(S+2)$ \\
$S+1$  & ${\sqrt{6} \over 42}   (2S-1)(S+6)$    &  $-{ 3\sqrt{70} \over 1225} (2S-3)(2S-1)(S-1)(2S+5)$ &
${\sqrt{3} \over 5} \sqrt{S(S+2)}$  & $- d^{121}_{000} {1 \over \sqrt{10}} (S-1)(2S-1)$ \\
$S+2$  & $-{\sqrt{6} \over 21}  (2S-1)S$       &  ${ 3\sqrt{70} \over 2450} (2S-3)(2S-1)(S-1)S$ &-&- \\
\end{tabular}
\end{ruledtabular}
\end{table*}

\section{Effective spin Hamiltonian*}
\label{sHeff}

Using the results of the previous section, a general equation can be derived for the second-order effective
spin Hamiltonian. A general procedure, which in principal leads to the same results as ours, has been
described recently by Liviotti {\it et al}.\cite{Car03} It consists of introducing new spin operators
$^{S-S'}\hat{M}^q_{k(e/c)}$, which in turn can be written as a combination of Stevens operators. However, no
simple and straightforward procedure is available to obtain the $\hat{M}$ operators. Furthermore, as the
analysis of experimental data is generally done in terms of the Stevens operators (or equivalents), one
almost always wants to have the $^{S-S'}\hat{M}^q_{k(e/c)}$ operators expressed by Stevens operators anyhow.
We believe that our approach has some advantages: it produces the effective spin Hamiltonian directly in
terms of ITOs (to which the Stevens operators are simply proportional), and, more importantly,
eq.~(\ref{Vexpand}) of the previous section provides a pretty easy way to find this decomposition.

The calculation of the first-order contribution to the effective spin Hamiltonian via eq.~(\ref{Texpand}) is
standard and needs no further comment.\cite{Ben90} The calculation of the second-order contributions is
enabled by noting that the perturbative part $\hat{H}_1$ of the microscopic spin Hamiltonian,
eq.~(\ref{Hito}), consists of a sum of spherical scalar products. $\hat{H}_1$ may thus be decomposed as
\begin{eqnarray}
 \hat{H}_1 &=& \sum_r \hat{H}_r,
 \cr\cr
 \hat{H}_r &=& \sum_{s_r}\sum_{q_r} T^{(k_r)*}_{q_r}({\bf V}_{s_r}) \hat{T}^{(k_r)}_{q_r}(s_r),
\end{eqnarray}
where $r$ (and $u$) indexes the individual terms in $H_1$. The second-order contribution produced by one
higher-lying spin multiplet is thus proportional to $\sum_{r,u} P_{\tau S} \hat{H}_{r} P_{\tau' S'}
\hat{H}_{u} P_{\tau S}$, see eq.~(\ref{Heff}). Inserting the individual terms of this sum in
eq.~(\ref{Vexpand}) leads to expressions like
\begin{eqnarray}
  \sum_k \Gamma_{k_r k}(s_r) \Gamma^*_{k_u k}(s_u) \: c_{k_r k_u k}
  \sum_{q_r q_u} \left(\begin{array}{ccc} k_r&k_u&k \\ -q_r&-q_u&q \end{array} \right)
  \cr\cr \times
  T^{(k_r)}_{-q_r}({\bf V}_{s_r}) T^{(k_u)}_{-q_u}({\bf V}_{s_u}) \hat{T}^{(k)}_q(S).
\end{eqnarray}
The sum over $q_r$ and $q_u$ is nothing else than the tensor product of two irreducible spherical tensors.
With that, the effective spin Hamiltonian up to second order may be conveniently written as
\begin{equation}
\label{Heff2nd}
 \hat{\bar{H}} = \sum_{r q_r} (-1)^{q_r} h^{1(k_r)}_{-q_r} \hat{T}^{(k_r)}_{q_r}(S)
 -{1 \over \Delta} \sum_{k q} (-1)^{q} h^{2(k)}_{-q} \hat{T}^{(k)}_q(S)
\end{equation}
with
\begin{eqnarray}
\label{Heff2nd2}
 h^{1(k_r)}_{q_r} &=& \sum_{s_r} \Gamma_{k_r}(s_r) T^{(k_r)}_{q_r}({\bf V}_{s_r}),
 \cr\cr
 h^{2(k)}_{q} &=& \sum_{r,u} \sum_{s_r,s_u} \Gamma_{k_r k}(s_r) \Gamma^*_{k_u k}(s_u)
 { (-1)^{k_r-k_u} c_{k_r k_u k} \over \sqrt{2k+1} } \cr\cr && \times
 [T^{(k_r)}({\bf V}_{s_r}) \otimes T^{(k_u)}({\bf V}_{s_u})]^{(k)}_q.
\end{eqnarray}
Here, only one higher-lying spin multiplet with a gap $\Delta$ has been considered, but the generalization to
more spin multiplets is obvious [only multiplets with $|S-S'| = 0, \ldots, min(k_r,k_u)$ contribute to
$h^{2(k)}_q$]. The similarity of the two terms in eq.~(\ref{Heff2nd}) with spherical scalar products is
satisfying to note. Unfortunately, such an elegant result cannot be obtained for the effective spin operators
$\hat{\bar{S}}_{i\alpha}$. Here terms like $P_{\tau S} \hat{H}_{r} P_{\tau' S'} \hat{T}^{(1)}_q(s) P_{\tau
S}$ appear in which only one sum over a component, $q_r$, is involved.

These equations shall be applied now to the case where the small part of the microscopic Hamiltonian consists
of only the bilinear anisotropic term, i.e., $\hat{H}_1 = \sum_{ij} \hat{{\bf S}}_i \cdot {\bf \Lambda}_{ij}
\cdot \hat{{\bf S}}_j$, were the tensors ${\bf \Lambda}_{ij}$ are assumed to be diagonal and traceless. This
situation is the one most often encountered in practice [in this case eqs.~(\ref{Heff2nd}) and
(\ref{Heff2nd2}) simplify: $k_r$ and $k_u$ are replaced by 2, $s_r$ and $s_u$ by $s$, and remaining indices
$r$,$u$ can be dropped]. The tensors ${\bf \Lambda}_s$ can be parameterized as usual by $D_s$ and $E_s$
parameters, and the non-zero components of the corresponding spherical tensors become $T^{(2)}_0({\bf
\Lambda}_s) = \sqrt{2/3} D_s$ and $T^{(2)}_2({\bf \Lambda}_s) = E_s$ (see also appendix~\ref{app_ITO}). After
a simple, but lengthy calculation the effective Hamiltonian eq.~(\ref{Heff2nd}) is obtained as
\begin{eqnarray}
\label{HeffG}
 \hat{\bar{H}} &=& D [\hat{S}^2_z - 1/3S(S+1)] + E (\hat{S}^2_x-\hat{S}^2_y)
 \cr\cr
  &&+ B^0_4 \hat{O}^0_4 + B^2_4 \hat{O}^2_4 + B^4_4 \hat{O}^4_4 ,
 \cr\cr
 D &=& \bar{D} - {\sqrt{2} \over \sqrt{3} \Delta} d^{222}_{000} [ (\bar{D}_2)^2 + 3 (\bar{E}_2)^2) ],
 \cr\cr
 E &=& \bar{E} + {\sqrt{2} \over \sqrt{3} \Delta} d^{222}_{000} ( \bar{D}_2 \bar{E}^*_2  +  \bar{D}^*_2 \bar{E}_2 ),
 \cr\cr
 B^0_4 &=& - {1 \over 6 \sqrt{70} \Delta} d^{224}_{000} [ 2(\bar{D}_4)^2 + (\bar{E}_4)^2) ],
 \cr\cr
 B^2_4 &=& - {\sqrt{5} \over 3 \sqrt{14} \Delta} d^{224}_{000} ( \bar{D}_4 \bar{E}^*_4  +  \bar{D}^*_4 \bar{E}_4 ),
 \cr\cr
 B^4_4 &=& - {\sqrt{35} \over 6 \sqrt{2} \Delta} d^{224}_{000} ( \bar{E}_4 )^2.
\end{eqnarray}
Here, we have expressed ITOs by the conventional operators using Table~\ref{tab2} to put the effective
Hamiltonian in the standard form (the constant $k=0$ term has been dropped). Also, we have introduced the
definition
\begin{eqnarray}
\label{Dk}
  \bar{D}_k = \sum_s D_s \Gamma_{2,k}(s),
\end{eqnarray}
where the index $k$ might be dropped, which means then $\bar{D} = \sum_s D_s \Gamma_2(s)$ (analogous
definitions were used for the $\bar{E}_k$ terms). It should be recalled that $s$ stands for the indices $ij$
or $i$, and that the sum over $s$ in the present case actually means $\sum_s = \sum_i + \sum_{i\neq j}$.
Again, only one higher-lying spin multiplet with a gap $\Delta$ has been considered. For the special case of
$S' = S-1$, eq.~(\ref{HeffG}) was derived already in Ref.~[\onlinecite{Car03}].

As expected, the mixing of spin multiplets due to the anisotropy leads to a renormalization of the
first-order zero-field-splitting parameters $\bar{D}$ and $\bar{E}$, and to the appearance of fourth-order
spin operators with strengths $B^0_4$, $B^2_4$, and $B^4_4$. In case of metal centers with spins larger than
3/2, the single-ion fourth-order spin terms in the microscopic Hamiltonian $\hat{H}$ also contribute to the
fourth-order terms in the effective spin Hamiltonian in first order. This adds terms $\bar{B}^m_4 = \sum_i
\Gamma_4(i) B^m_4(i)$ to the $B^m_4$'s in eq.~(\ref{HeffG}).

%

\section{Inelastic neutron scattering}
\label{sINS}

In this section, we analyze the effects of spin mixing on the INS intensity of a powder sample in zero
magnetic field. A discussion of the situation encountered in EPR, which additionally requires the
consideration of a magnetic field, is deferred to the next section. The INS intensity in a spin cluster is
given by the following cross section
\begin{equation}
\label{cross_section}
 {d^2 \sigma \over d\Omega d\omega} =
 C(Q,T) \sum_{nm} {e^{-\beta E_n} \over Z(T)} I_{nm}({\bf Q}) \delta(\omega-{E_m-E_n\over \hbar}),
\end{equation}
were $C(Q,T)=(\gamma e^2/m_e c^2) (k'/k) \exp[-2 W(Q,T)]$, $\beta = 1/(k_B T)$, and $Z(T)$ is the partition
function.\cite{Fur79,Gue85} The sum runs over all transitions $|n\rangle \rightarrow |m\rangle$. For a
powder sample in zero magnetic field $I_ {nm}$ is given by
%
\begin{eqnarray} \label{Inm}
 I_{nm}(Q) &=& \sum_{ij} F^*_i(Q) F_j(Q) \{ {2 \over 3} j_0(Q R_{ij}) {\bf \tilde{S}}_i \cdot {\bf \tilde{S}}_j
 \cr && +
 j_2(Q R_{ij}) \sum_q T^{(2)*}_q({\bf R}_{ij})
 \cr && \times
 \left[T^{(1)}({\bf \tilde{S}}_i) \otimes T^{(1)}({\bf \tilde{S}}_j)\right]^{(2)}_q \}.
\end{eqnarray}
%
In this equation, $F_i(Q)$ is the magnetic form factor of the $i$th spin center, $j_k$ is the spherical
Bessel function of order $k$, and ${\bf R}_{ij} = {\bf R}_{i} - {\bf R}_{j}$ is the distance vector between
the $i$th and $j$th ion [explicit expressions of eq.~(\ref{Inm}) are provided in appendix~\ref{app_INS}].
The ordered products $\tilde{S}_{i \alpha} \tilde{S}_{j \beta}$, which appear in eq.~(\ref{Inm}), stand for
$ \langle n|\hat{S}_{i \alpha}|m\rangle\langle m| \hat{S}_{j \beta}|n\rangle$, where $|n\rangle$ denotes an
eigenstate of the Hamiltonian $\hat{H}$.

For the evaluation of the INS intensity of the transitions within a spin multiplet, one inserts $\langle
n|\hat{S}_{i \alpha}|m\rangle = \langle \bar{n}|\hat{\bar{S}}_{i \alpha}|\bar{m}\rangle$ into
eq.~(\ref{Inm}), determines the effective spin operators $\hat{\bar{S}}_{i\alpha}$, and then substitutes them
by single-spin operators $\hat{S}_{\alpha}$. Unfortunately, no compact result is available for the effective
spin operators, and accordingly also not for the INS intensity. The calculations are pretty straightforward,
but lengthy, as are the results. Thus, in the next subsection (which may be skipped by readers not interested
in the details), we limit our calculation of the $\hat{\bar{S}}_{i\alpha}$ to the case of a bilinear
anisotropy described by diagonal and traceless tensors ${\bf \Lambda}_{ij}$. It is, however, general enough
to account for most situations met in practice. In order to gain a feeling for the effects to be expected, in
particular as compared to the single-spin approach, we then - in a second subsection - consider shortly the
first-order effects and then specialize to the case of a uniaxial anisotropy, for which we work out the INS
intensity up to second order analytically. In a third subsection we finally discuss three examples
exhibiting uniaxial magnetic anisotropy. Since in this work we are interested in transitions within an
individual spin multiplet, an uniaxial system is the minimal model of relevance (concerning transitions
between different spin multiplets, we mention e.g. Ref.~\onlinecite{OW_INS}).

\subsection{Effective spin operators*}
\label{sINSA}

The effective spin operators $\hat{\bar{S}}_{i \alpha}$ within a $\tau S$ subspace shall be calculated for a
system described by $\hat{H} = \hat{H}_0 + \hat{H}_1$ with $\hat{H}_1 = \sum_{ij} \hat{{\bf S}}_i \cdot {\bf
\Lambda}_{ij} \cdot \hat{{\bf S}}_j$, where the ${\bf \Lambda}_{ij}$ are assumed to be diagonal and traceless
as in section~\ref{sHeff}. According to eq.~(\ref{Oeff}), the calculation of the second-order contribution
involves the terms $P_{\tau S} \hat{T}^{(2)}_{q_2}(s) P_{\tau' S'} \hat{T}^{(1)}_{q_1}(i) P_{\tau S}$ +
$P_{\tau S} \hat{T}^{(1)}_{q_1}(i) P_{\tau' S'} \hat{T}^{(2)}_{q_2}(s) P_{\tau S}$. In employing
eq.~(\ref{Vexpand}), the values of $k$ are restricted to $k= 1,2,3$. Because in our case of an isotropic
zero-order Hamiltonian $\hat{H}_0$ the reduced matrix elements may be chosen as real, and because of
$d^{21k}_{q_1 q_2 q} = (-1)^{k+1} d^{12k}_{q_2 q_1 q}$, the $k=2$ contribution cancels out (we nevertheless
keep complex writing since sometimes it has advantages to work with complex wave functions, e.g., when
exploiting spatial symmetries \cite{OW_SYM,OW_INS}). Re-expressing the coefficients $d^{12k}_{q_1 q_2 q}$ by
$d^{12k}_{000}$, the effective spin operators finally become
\begin{widetext}
\begin{eqnarray}
\label{Seff}
  \hat{\bar{S}}_{i{x \over y}} &=&
  \left\{ \Gamma_1(i) + {d^{121}_{000} [ \bar{\bar{D}}_1(i) \mp 3 \bar{\bar{E}}_1(i) ]  \over \sqrt{6} \Delta}
  - { d^{123}_{000} [ 2 \bar{\bar{D}}_3(i) \mp \bar{\bar{E}}_3(i)]  \over 2 \sqrt{15} \Delta} [5\hat{S}_z^2 - S(S+1)-5\hat{S}_z+2]
  - { 5 d^{123}_{000} \bar{\bar{E}}_3(i) \over 2 \sqrt{15} \Delta} ( \hat{S}^2_x - \hat{S}^2_y )
  \right\} \hat{S}_{x \over y}
  \cr
  &&\pm { 5 d^{123}_{000} \bar{\bar{E}}_3(i) \over 2 \sqrt{15} \Delta} ( \hat{S}_x \hat{S}_y + \hat{S}_y \hat{S}_x ) \hat{S}_{y \over x},
 \cr\cr
  \hat{\bar{S}}_{iz} &=& \left\{ \Gamma_1(i) - {2 d^{121}_{000} \bar{\bar{D}}_{1}(i) \over \sqrt{6} \Delta}
  - { d^{123}_{000} \bar{\bar{D}}_3(i) \over \sqrt{15} \Delta} [5\hat{S}_z^2 - 3S(S+1)+1] \right\} \hat{S}_z
  - { 5 d^{123}_{000} \bar{\bar{E}}_3(i) \over \sqrt{15} \Delta} (\hat{S}_z-1)( \hat{S}_x^2 - \hat{S}_y^2
  ).
\end{eqnarray}
\end{widetext}
Here again, only one higher-lying spin multiplet at energy $\Delta$ was assumed (only states with $|S-S'| =
0,1$ may contribute). Also, we have introduced the abbreviation
\begin{eqnarray}
\label{Dhatk}
 \bar{\bar{D}}_k(i) = \bar{D}_k \Gamma_{1,k}^*(i) + \bar{D}_k^* \Gamma_{1,k}(i)
\end{eqnarray}
with $\bar{D}_k$ defined as in section~\ref{sHeff}, eq.~(\ref{Dk}) (and analogous definitions for the
$\bar{\bar{E}}_k$ terms).

The various terms appearing in $\hat{\bar{S}}_{i \alpha}$ shall be analyzed. In each of the above
expressions, the first terms, which are the first-order contributions, are constants within a $\tau S$
subspace (or $|SM\rangle$ space, respectively) and equal for all components $\alpha = x,y,z$. The second
terms are also constants within the $|SM\rangle$ space, but introduce an anisotropy in the intensities as
their signs and magnitude are different for the $x$, $y$, and $z$ components. The remaining terms, finally,
depend via the $\hat{S}_\alpha$-dependent prefactors explicitly on the particular states involved in a
transition, and thus lead to differences in the relative intensities of the transitions within a spin
multiplet.

In principle, if ions with spins larger than 3/2 are involved, there would be also contributions from the
fourth-order single-ion terms in the microscopic spin Hamiltonian. However, in contrast to the situation for
the effective spin Hamiltonian where these terms had to be included (though in first order was sufficient),
they can be safely neglected here, because they appear in second order as the bilinear terms, but are
significantly weaker than those.

\subsection{INS intensity}
\label{sINSB}

We are now prepared to discuss the intensity of the INS transitions within a spin multiplet. First, we
consider the scattering intensity up to first-order only, as this allows a general result. In this
approximation, the effective spin operators are simply given by $\hat{\bar{S}}_{i\alpha} = \Gamma_1(i)
\hat{S}_\alpha$ [the projection coefficient $\Gamma_1(i)$ is defined in eq.~(\ref{Texpand})], which after
insertion into the INS formula eq.~(\ref{Inm}) yields
\begin{subequations}
\begin{eqnarray}
 \label{I1sta}
 I^{1st}_{n m}(Q) &=& {2 \over 3} (-\sqrt{3}) f^{(0)}_{0}(Q) {\bf \tilde{S}} \cdot {\bf \tilde{S}}
 \cr
 &&+ \sum_q f^{(2)}_{q}(Q) \left[T^{(1)}({\bf \tilde{S}}) \otimes T^{(1)}({\bf \tilde{S}})\right]^{(2)}_q
 ,\qquad
 \\
  f^{(k)}_{q}(Q) &=& \sum_{ij} F^*_i(Q) F_j(Q) j_k(Q R_{ij})
 \cr
  &&\times  T^{(k)*}_q({\bf R}_{ij}) \Gamma_1(i) \Gamma_1(j),
\end{eqnarray}
\label{I1st}
\end{subequations}
where the factor $-\sqrt{3}$ in the first term on the r.h.s. of eq.~(\ref{I1sta}) compensates for $T^{(0)}_0$
in the definition of $f^{(0)}_{0}$ [since $ T^{(0)}_0({\bf R}_{ij}) = - 1/ \sqrt{3}$]. The functions
$f^{(k)}_{q}(Q)$ may be considered as the (first-order) interference factors appropriate for the problem, as
they produce a $Q$ dependence. Appearing in first-order, they are actually the main source for the $Q$
dependence of the INS intensity.

The first-order result eq.~(\ref{I1st}) is interesting because it clearly demonstrates the deficits of the
single-spin approach. Only the first term on the r.h.s. of eq.~(\ref{I1sta}) is retained in the single-spin
approach, but with the interference factor $f^{(0)}_{0}(Q)$ omitted (it may be noted that $2 {\bf \tilde{S}}
\cdot {\bf \tilde{S}} = 2 |\langle \bar{n}|\hat{\bar{S}}_{z}|\bar{m}\rangle|^2 + |\langle \bar{n}|
\hat{\bar{S}}_{+}|\bar{m}\rangle|^2 + |\langle \bar{n}| \hat{\bar{S}}_{-}|\bar{m}\rangle|^2$). The effects of
neglecting the $Q$ dependence were discussed already in section~\ref{sSSApproach}. It should be noted that
for transitions within a spin multiplet, the $k=2$ terms in eq.~(\ref{I1st}) are always of importance so that
the $Q$ dependence missed in the single-spin approach is not only that of $f^{(0)}_{0}(Q)$ [an example is
given below through eq.~(\ref{Iuni})].

Equation~(\ref{I1st}) becomes quite useful when the $\Gamma_1(i)$ are known, e.g., from approximative wave
functions or numerical calculations, as it then allows to calculate the INS intensity correctly up to
first-order, which often is a great improvement as compared to the single-spin approach. A further
interesting possibility emerges. In general, the projection coefficients $\Gamma_1(i)$ are affected by the
exchange coupling topology as employed in $\hat{H}_0 = - J_{ij} \sum_{ij} \hat{{\bf S}}_i \cdot \hat{{\bf
S}}_j$. An analysis of the $Q$ dependence in principle thus provides information about the exchange coupling
constants in the system under consideration.

We proceed with the calculation of the INS intensity up to second order for a system with uniaxial magnetic
anisotropy, i.e., a system described by the microscopic Hamiltonian $\hat{H} = \hat{H}_0 + \sum_{ij}
\hat{{\bf S}}_i \cdot {\bf \Lambda}_{ij} \cdot \hat{{\bf S}}_j$, where the ${\bf \Lambda}_{ij}$ are diagonal
and fulfill $\Lambda_{xx} = \Lambda_{yy} = - \Lambda_{zz}/2$. The second-order effective spin Hamiltonian is
then $\hat{\bar{H}} = D[ \hat{S}^2_z - 1/3S(S+1) ] + B^0_4 \hat{O}^0_4$ with parameters as given in
eq.~(\ref{HeffG}). Furthermore, the eigenstates of $\hat{\bar{H}}$ are simply the $|SM\rangle$ states, and
the calculation of matrix elements is elementary (in general $|\bar{n}\rangle$ will be a linear combination
of $|SM\rangle$ spin states with different magnetic quantum numbers $M$).

As only transitions with $M \rightarrow M+1$ (and $M+1 \rightarrow M$) are relevant, the $\alpha = z$ terms
drop out of the formula for the INS intensity, eq.~(\ref{Inm}). One is left with considering
$\hat{\bar{S}}_{ix}$ and $\hat{\bar{S}}_{iy}$. For a uniaxial system their structure is of the form
$\hat{\bar{S}}_{i,x/y} = [a_i + b_i - c_i F(\hat{S}_z)] \hat{S}_{x/y}$, with site-dependent constants $a_i$,
$b_i$, $c_i$, and $F(\hat{S}_z) = 5\hat{S}_z^2 - S(S+1)-5\hat{S}_z+2$, see eq.~(\ref{Seff}). The matrix
element $\tilde{S}_{ix}\tilde{S}_{jx} = \langle M | \hat{\bar{S}}_{ix} | M+1 \rangle \langle M+1 |
\hat{\bar{S}}_{jx} | M \rangle$ equals, up to first order in the gap $\Delta$, $[ a_i a_j + a_i b_j + b_i a_j
- a_i c_j F(M+1) - c_i a_j F(M) ] \tilde{S}_{x} \tilde{S}_{x}$, and similarly for $\tilde{S}_{iy}
\tilde{S}_{jy}$. Because of the sum $\sum_{ij}$ in the INS intensity formula, and because of $\sum_{ij} a_i
b_j = \sum_{ij} b_i a_j$ and similarly for $a_i c_j$, one may simplify to $\{ a_i a_j + 2 a_i b_j - a_i c_j
[F(M+1) + F(M) ]\} \tilde{S}_{x} \tilde{S}_{x}$. With these and similar results for $\tilde{S}_{iy}
\tilde{S}_{jy}$, with $F(M+1)+F(M) = 2[5M^2-S(S+1)+2]$, $\tilde{S}_{x} \tilde{S}_{x} = \tilde{S}_{y}
\tilde{S}_{y} = [S(S+1)-M(M+1)]/4$, and the explicit expressions for the constants $a_i$, $b_i$, $c_i$
[eq.~(\ref{Seff})], one finally obtains
\begin{eqnarray}
\label{Iuni}
  I^{uni}_{M,M+1}(Q) &=& {1 \over 3} f_{M}(Q) \left[ S(S+1)-M(M+1) \right],
  \cr\cr
  f_{M}(Q) &=&  \bar{f}(Q) + {4 d^{121}_{000} \bar{D}_1 \over \sqrt{6} \Delta } \bar{f}_1(Q)
  \cr && -
  {4 d^{123}_{000} \bar{D}_3 \over \sqrt{15} \Delta } \left[5M^2 -S(S+1)+2 \right] \bar{f}_3(Q),
 \cr\cr
  \bar{f}_k(Q) &=& \sum_{ij} F^*_i(Q) F_j(Q) [ j_0(Q R_{ij})
  \cr && -
  {1 \over 2}  j_2(Q R_{ij})  C^2_0({\bf R}_{ij}) ] \: \Gamma_1(i) \Gamma_{1,k}(j).
\end{eqnarray}
Here, $I^{uni}_{M+1,M} = I^{uni}_{M,M+1}$ and $C^2_0({\bf R}_{ij}) = [3(R_{ij,z}/R_{ij})^2 - 1]/2$. In the
definition of the terms $\bar{f}_k(Q)$, the index $k$ might be dropped, which means to use $\Gamma_{1}(j)$
instead of $\Gamma_{1,k}(j)$. The coefficients $d^{121}_{000}$ and $d^{123}_{000}$ are given in
Tab.~\ref{tab1}, $\bar{D}_1$ and $\bar{D}_3$ are defined in eq.~(\ref{Dk}), and the projection coefficients
$\Gamma_{1,k}(i)$ in eq.~(\ref{d}). We have assumed $\bar{D}_k$ and $\Gamma_{1,k}(i)$ to be real for
simplicity. It is useful to note that because of the properties of the projection coefficients
$\Gamma_{1,k}(i)$ the relation $\bar{D}_3 \bar{f}_3 = (\langle S \| \hat{T}^{(1)} \| S \rangle / \langle S \|
\hat{T}^{(3)} \| S \rangle )^2 \bar{D}_1 \bar{f}_1$ holds, i.e., that up to a scaling factor $\bar{f}_1(Q)$
and $\bar{f}_3(Q)$ are the same.

The function $f_{M}(Q)$ may be again considered as the appropriate interference factor, and eq.~(\ref{Iuni})
demonstrates all the effects which we wish to point out in this work. The first term in $f_{M}(Q)$,
$\bar{f}(Q)$, corresponds exactly to the first-order contribution calculated in eq.~(\ref{I1st})
[$\bar{f}(Q) = -\sqrt{3} f^{(0)}_0(Q) - 3/2 f^{(2)}_0$(Q)]. As expected, it leads to a $Q$ dependence which
is equivalent for all transitions within a spin multiplet, i.e., is independent on $M$. The spin mixing
manifests itself in two ways. First, it adds to the overall $Q$ dependence via the second term in $f_M(Q)$.
This contribution is independent of $M$ like the first-order contribution, but smaller by a factor of $1 /
\Delta$. Thus, from an experimental point of view, it is a minor correction which we expect is difficult to
be separated experimentally from $\bar{f}(Q)$. The second effect of spin mixing, however, brought in by the
third term, leads to different $Q$ dependencies of the individual transitions because of its explicit
dependence on $M$. This effect should be amenable to experimental detection as a more sensitive differential
data analysis can be applied. For instance, one could analyze the differences
$I^{uni}_{M,M+1}/[S(S+1)-M(M+1)] - I^{uni}_{M',M'+1}/[S(S+1)-M'(M'+1)]$.

The second-order projection coefficients $\Gamma_{1,k}(j)$ are proportional to the reduced matrix elements
$\langle \tau S \| \hat{T}^{(1)}(j) \| \tau' S' \rangle$ [eq.~(\ref{d})], and thereby also to what we called
the spin transition map. Thus $\bar{f}_1(Q)$ and $\bar{f}_3(Q)$, respectively, arise from the correlation of
the geometrical structure with the spin transition map. Equation~(\ref{Iuni}) confirms the intuitive
expectation that a mixing with higher lying spin multiplets, which mixes in spin transition maps with
different topologies than that of the ground-state multiplet, leads to different $Q$ dependencies for each
transition within a ground-state multiplet.

\subsection{Examples}
\label{sINSC}

In this section, three examples of  molecular spin clusters, a fictitious but instructive hetero-nuclear
dimer, the Mn-[3~$\times$~3] grid, and the Mn$_{12}$ cluster, will be discussed in order to underpin the
above considerations.

\begin{figure}
\includegraphics{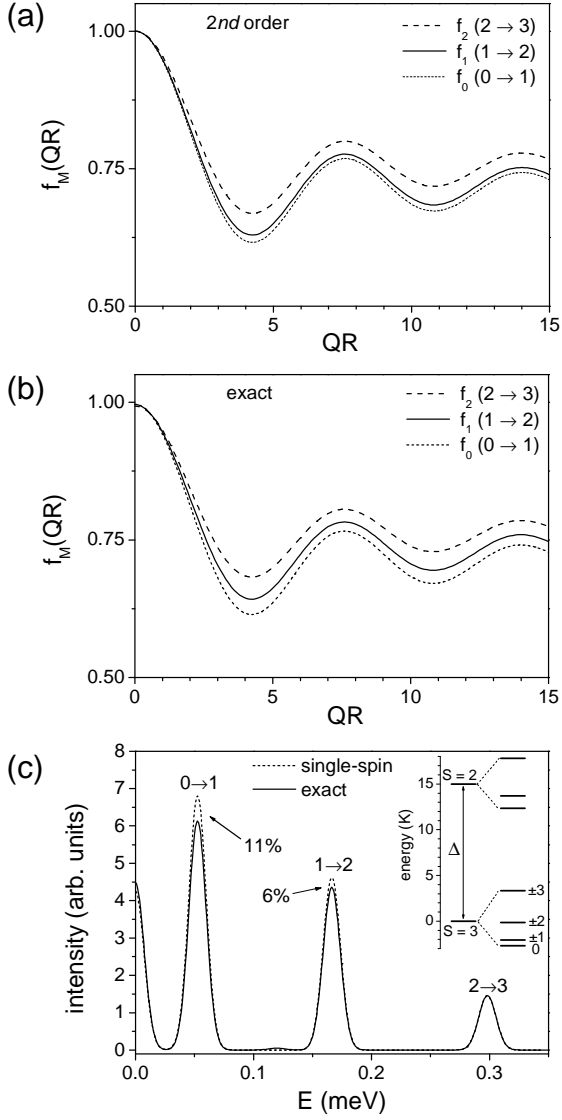}
\caption{\label{fig1} Results for the ferromagnetic spin-1/2-spin-3/2 dimer with uniaxial anisotropy
considered in the text. (a) and (b) show the $Q$ dependence of the interference factors $f_{M}$ for the three
transitions $0 \rightarrow 1$, $1 \rightarrow 2$, and $2 \rightarrow 3$ within the $S = 3$ multiplet as
calculated by second-order perturbation theory, eq.~(\ref{Iuni}), and numerically from the microscopic spin
Hamiltonian, respectively. (c) INS spectrum as calculated numerically from the microscopic Hamiltonian and
the single-spin approach for a temperature $T$ = 3~K, momentum transfer $QR$ = 4, and Gaussian lines with a
FWHM of 20~$\mu$eV. The curves were scaled so as to coincide for the $2 \rightarrow 3$ peak at 0.3~meV. The
inset shows the energy spectrum of the dimer.}
\end{figure}

As a first example a ferromagnetic spin-5/2 - spin-1/2 dimer with uniaxial anisotropy is considered. This
example may seem somewhat artificial, but many hetero-nuclear dimers exist, and dimer clusters may evolve
into model systems for the study of the many-spin effects on the INS intensity as the analysis becomes
particularly straightforward. The microscopic spin Hamiltonian is assumed as $\hat{H} = -J \hat{{\bf S}}_1
\cdot \hat{{\bf S}}_2 + D_1 \hat{S}_{1,z}^2$ with $S_1 = 5/2$ and $S_2 = 1/2$. The energy spectrum is
simple, it consists of a $S = 3$ ground-state multiplet and a higher lying $S = 2$ multiplet at energy
$\Delta = 3 J$, which both are zero-field split due to the anisotropy. In the following, the values $J =
5$~K and $D_1 = 1$~K will be assumed. For these parameters, the energy spectrum consists of well separated
spin multiplets as shown in the inset of Fig.~\ref{fig1}(c).

The second-order effective spin Hamiltonian of the $S = 3$ ground-state multiplet is $\hat{\bar{H}} = D[
\hat{S}^2_z - 1/3S(S+1) ] + B^0_4 \hat{O}^0_4$ [as obtained from eq.~(\ref{HeffG}) for a uniaxial cluster].
For the calculation of the parameters $D$ and $B^0_4$, the first-order projection coefficient $\Gamma_2(1)$
and the second-order projection coefficients $\Gamma_{2,2}(1)$ and $\Gamma_{2,4}(1)$, which enter $\bar{D}$,
$\bar{D}_1$ and $\bar{D}_3$ [eqs.~(\ref{HeffG}) and (\ref{Dk})], need to be evaluated, or the two reduced
matrix elements $\langle 5/2, 1/2, 3 \| \hat{T}^2(S_1) \| 5/2, 1/2, 3 \rangle$ and $\langle 5/2, 1/2, 3 \|
\hat{T}^2(S_1) \| 5/2, 1/2, 2 \rangle$, respectively [see eqs.~(\ref{Texpand}) and (\ref{d})]. Here, we have
written the zero-order dimer states as $|S_1 S_2 S M\rangle$. The respective coefficients for $S_2$ are zero
as a spin-1/2 cannot exhibit a zero-field-splitting. Using standard ITO methods, the calculation is easy
enough and one obtains $D$ = $2/3 D_1 + 4/63 D_1^2/\Delta$ = 0.6709~K and $B^0_4$ = $1/315 D_1^2/\Delta$ =
0.2116~mK [the other parameters in eq.~(\ref{HeffG}) are obviously zero]. If one fits the exact eigenvalues
of the states of the $S = 3$ multiplet to the effective spin Hamiltonian $\hat{\bar{H}}$, one obtains $D$ =
0.6707~K and $B^0_4$ = 0.1770~mK, showing that the analytic second-order results give the corrections $D -
2/3 D_1$ and $B^0_4$ correctly up to 5\% and 16\%, respectively.

Next, the INS intensity, eq.~(\ref{Iuni}), will be evaluated. The parameters $\bar{D}_1$ and $\bar{D}_3$,
respectively, have been already obtained in the above calculation of the effective spin Hamiltonian. For the
further evaluation it is useful to note the special properties of a dimer, namely that $\Gamma_1(1) +
\Gamma_1(2) = 1$, $\Gamma_1(1)/\Gamma_1(2) = S_1/S_2$, and $\Gamma_{1,k}(2) = - \Gamma_{1,k}(1)$. The first
two relations determine $\Gamma_1(1) = 5/6$ and $\Gamma_1(2) = 1/6$, which yields $\bar{f}(Q) = 1 - 5/18
\mathcal{F}(Q)$ with $\mathcal{F}(Q) = 1 - j_0(QR) + 1/2 j_2(QR)$, and the third relation yields
$\bar{f}_k(Q) = 2/3 \Gamma_{1,k}(1) \mathcal{F}(Q)$. One is thus left with the calculation of the reduced
matrix element $\langle 5/2, 1/2, 3 \| \hat{T}^1(S_1) \| 5/2, 1/2, 2 \rangle$ entering $\Gamma_{1,k}(1)$,
which is again easy enough. One finally obtains the result $f_M(Q) = 1 - [ 5/18 - 4/27 D_1^2/\Delta (M^2-1)]
\mathcal{F}(Q)$, which is plotted in Fig.~\ref{fig1}(c) for the three possible transitions within the $S = 3$
multiplet. For comparison, the result of an exact numerical calculation is shown in Fig.~\ref{fig1}(b). The
agreement of the second-order and exact results is good. In particular, the correct trend and order of
magnitude is recovered, though the second-order results for $f_1(Q)$, which corresponds to the transition $1
\rightarrow 2$, lie significantly too low at larger $Q$ values [for $QR \gg 1$ the $f_1(Q)$ curve in
Fig.~\ref{fig1}(a) is visibly shifted towards the $f_0(Q)$ curve as compared to Fig.~\ref{fig1}(b)].

These results demonstrate the considerations of the previous subsection, in particular the dependence on $M$
is apparent. In the case of a dimer, the first- and second-order contributions exhibit identical $Q$
dependencies, i.e., $\bar{f}(Q)$ and $\bar{f}_k(Q)$ in eq.~(\ref{Iuni}) are both proportional to the same
function $\mathcal{F}(Q)$, since the spatial symmetry of a dimer completely fixes the functional dependence
on $Q$.\cite{OW_INS} In the general case, however, $\bar{f}(Q)$ and $\bar{f}_k(Q)$ will exhibit different
$Q$ dependencies because of the different topologies of the spin transition maps involved in $\bar{f}(Q)$ and
$\bar{f}_k(Q)$, respectively.

The dependence of the interference factors on $M$ obviously leads to changes in the intensity pattern of the
transitions as compared to that calculated in the single-spin approach, or in first order, respectively (both
the single-spin approach and the first-order result disregard the $M$ dependence and are similar in this
regard). This is shown in Fig.~\ref{fig1}(c), which presents the neutron energy-loss spectrum as calculated
exactly and in the single-spin approach (the results were obtained numerically). The calculated intensities
were normalized so as to coincide for the $2 \rightarrow 3$ transition at 0.3~meV. Clearly, the many-spin
effects lead to a significant effect on the INS intensity, which is on the order of 10\% for the present
system. This is encouraging, as it suggests that such effects indeed could be observable. It is also
noteworthy that the many-spin effects on the INS intensity are much stronger than on the
zero-field-splitting parameter $D$, which is affected by only 0.6\%.

The above example allows to elucidate another important point. If the spin-5/2 center would additionally
exhibit a single-ion forth-order anisotropy, which would add a term $B_1 \hat{O}^0_4(S_1)$ to the microscopic
spin Hamiltonian, the fourth-order term in the effective spin Hamiltonian would become $B^0_4 = 1/315
D_1^2/\Delta + 1/3 B_1$. Clearly, an analysis of the energy spectrum or transition energies alone cannot
distinguish between the second- and first-order contributions to $B^0_4$ coming from $D_1$ and $B_1$,
respectively. In contrast, the many-spin effects on the INS intensity are governed by the spin mixing due to
the $D_1$ term, while the $B_1$ term is ineffective (see the discussion at the end of section~\ref{sINS}.A).
In other words, in the INS spectrum, Fig.~\ref{fig1}(c), the transition energies would shift with an
additional $B_1$ term, but the intensities would not be affected. Thus, a simultaneous analysis of energy
spectrum and transition intensities allows, in principle, to separate the parts in the cluster fourth-order
term which come from spin mixing and from single-ion anisotropy. This has great potential. With the advent of
the single-molecule magnets the understanding of the origin of magnetic anisotropy, and its eventual
controlling, has grown enormously in importance. Fourth-order terms in the effective spin Hamiltonian such
as a $B^4_4$ term are of particular relevance because of their crucial role in the process of quantum
tunneling of the magnetization.\cite{Mir99,Mn12_Fe8} INS studies as the one just suggested have the
potential to evolve into a powerful tool in this regard.

As a second example, the so called Mn-[3~$\times$~3] grid will be considered. This system is of some
relevance for the present work as it provided a clear-cut demonstration of the importance of spin mixing via
the observation of magneto-oscillations in the torque signal.\cite{OW_QMO} In the Mn-[3~$\times$~3] grid
molecule, nine spin-5/2 Mn(II) ions occupy the positions of a regular 3~$\times$~3 matrix, held in place by
a lattice of organic ligands. The parameters of the microscopic spin Hamiltonian were determined from
magnetization, high-field torque, and INS studies.\cite{OW_Mn3x3,OW_QMO,Gui04} The Mn ions within a molecule
experience an antiferromagnetic interaction leading to a $S$ = 5/2 ground state, which exhibits an easy-axis
zero-field-splitting with $D = -0.48$~K. In the following, the two transitions $1/2 \rightarrow 3/2$ and
$3/2 \rightarrow 5/2$ within the $S = 5/2$ ground-state multiplet are the subject of interest. They have
been observed by INS, though with poor statistics.\cite{Gui04}

\begin{figure}
\includegraphics{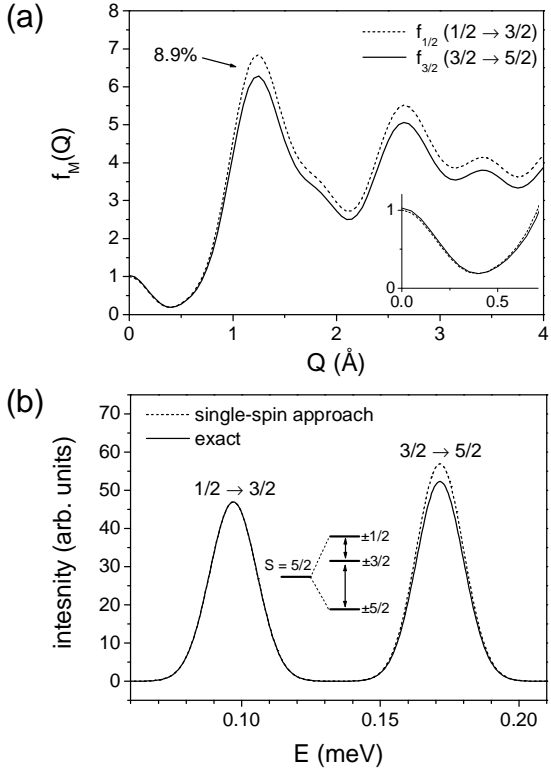}
\caption{\label{fig2} Results of numerical calculations for the Mn-[3~$\times$~3] cluster. (a) $Q$ dependence
of the interference factors $f_{M}$ for the two transitions $1/2 \rightarrow 3/2$ and $3/2 \rightarrow 5/2$
within the $S = 5/2$ ground state multiplet as calculated from the microscopic spin Hamiltonian. The inset
shows the low-$Q$ range with greater detail. (b) INS spectrum as calculated from the microscopic Hamiltonian
and the single-spin approach for a temperature $T$ = 3~K, momentum transfer $Q$ = 1.5~{\AA}, and Gaussian
lines with a FWHM of 20~$\mu$eV. The curves were scaled so as to coincide for the $1/2 \rightarrow 3/2$ peak
at 0.097~meV. The inset shows a sketch of the zero-field-split $S = 5/2$ ground-state multiplet of
Mn-[3~$\times$~3].}
\end{figure}

Using sparse-matrix techniques,\cite{sparse} the interference factors $f_{1/2}(Q)$ and $f_{3/2}(Q)$
characterizing the two transitions under consideration as well as the neutron energy-loss spectrum were
calculated numerically from the full microscopic spin Hamiltonian [for the Hamiltonian see eq.~(1) of
Ref.~\onlinecite{OW_QMO} and the parameters given there]. The results are presented in Fig.~\ref{fig2}. They
once more demonstrate all the effects of spin mixing identified before. In particular, the interference
factors shown in Fig.~\ref{fig2}(a) are again clearly different for the two transitions, i.e., are dependent
on $M$. The exact intensities in fact differ by as much as 9\% as compared to the expectation from the
single-spin approach. The peak intensities in the neutron scattering spectra are accordingly affected too, as
demonstrated in Fig.~\ref{fig2}(b). This example thus shows, that the effects of spin mixing on the INS
intensities indeed can be of significant size in experimentally relevant systems. First hints for the
relevance of such effects have been in fact found in a previous INS study.\cite{ Gui04}

The example of the Mn-[3~$\times$~3] grid also allows to demonstrate the other effect of spin mixing, which
was not displayed by the above dimer because of its particular symmetry properties. In Mn-[3~$\times$~3],
the $Q$ dependencies of the two transitions differ not only in magnitude, but also exhibit different
functional dependencies. This is evidenced in the inset of Fig.~\ref{fig2}(a), which shows the behavior of
the $Q$ dependence of the two transitions $1/2 \rightarrow 3/2$ and $3/2 \rightarrow 5/2$ at low momentum
transfer in greater detail. A crossing of the two curves is evident, which indicates that the functions
$\bar{f}(Q)$ and $\bar{f}_k(Q)$ in eq.~(\ref{Iuni}) exhibit different functional dependencies on $Q$ because
of the different spin transition maps involved in $\bar{f}(Q)$ and $\bar{f}_k(Q)$.

As a final example, the famous Mn$_{12}$ cluster will be considered.\cite{Mn12_Fe8} This single-molecule
magnet consists of a tetrahedral core of four spin-3/2 Mn(IV) ions surrounded by a crown of eight spin-2
Mn(III) ions, and is characterized by a $S = 10$ ground state with a large easy-axis zero-field-splitting
($D$ = -0.66~K, $B^0_4$ = -34~$\mu$K, $B_4^4 = \pm 43~\mu$K).\cite{Mir99,Bir04} Furthermore, starting at
about 40~K a number of higher lying spin multiplets with $S$ = 9, 10, and 11 exist, which are so low in
energy that the lowest $S = 9$ multiplet actually overlaps with states of the zero-field-split $S = 10$
ground-state multiplet [see Fig.~7(a) in Ref.~\onlinecite{Cha04}]. Because of this, one expects pronounced
spin-mixing effects in this cluster.

\begin{figure}
\includegraphics{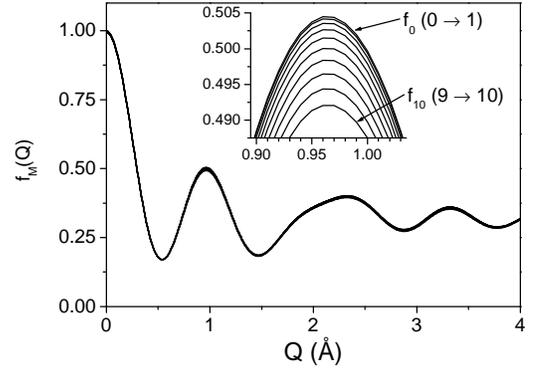}
\caption{\label{fig3} Numerically calculated $Q$ dependence of the interference factors $f_{0}$ to $f_{9}$
for the ten transitions within the $S$ = 10 ground state of the Mn$_{12}$ cluster (for details see text). The
curves essentially fall onto each other. The inset provides a closer view on the peak at about $Q$ =
1~{\AA}. The individual curves can be distinguished, but the differences are very small.}
\end{figure}

Using again sparse-matrix techniques,\cite{sparse} the energies of the states with $M = 0, \ldots, 10$ of the
$S = 10$ ground-state multiplet, as well as the interference factors $f_{0}(Q)$ to $f_{9}(Q)$ for the
respective transitions between them, were calculated numerically from the microscopic spin Hamiltonian. The
Hamiltonian consisted of the isotropic exchange terms appended by uniaxial single-ion anisotropy terms. The
exchange part and the respective coupling constants were chosen as in Ref.~\onlinecite{Cha04}. Since the
Mn(III) Jahn-Teller ions are expected to be the main source of magnetic anisotropy, the single-ion anisotropy
was modelled by a term $D_{Mn(III)} \sum S_{i,z}^2$, where the sum runs over the eight Mn(III) ions. From the
first-order projection coefficients one estimates $D = 0.2014 D_{Mn(III)}$, and $D_{Mn(III)}$ = -3.3~K was
chosen for the calculations. The obtained interference factors are plotted in Fig.~\ref{fig3}. The ten curves
are basically indistinguishable and coincide with the first-order contribution $\bar{f}(Q)$ [which is easy to
calculate analytically from eq.~(\ref{I1st}) thanks to the tables given in Ref.~\onlinecite{Cha04}]. Thus,
surprisingly, spin mixing is very weak in Mn$_{12}$. This finding is further supported by fitting the
calculated energies of the states of the $S$ = 10 ground state to the effective spin Hamiltonian
$\hat{\bar{H}} = D[ \hat{S}^2_z - 1/3S(S+1) ] + B^0_4 \hat{O}^0_4$, which yields $D$ = -0.653~K and $B^0_4$
= -1.97~$\mu$K ($B^4_4$ is obviously zero in our model), i.e., yields a value for $B^0_4$ which is one order
of magnitude smaller than the experimental one (but at least of the correct sign).

The absence of spin mixing in Mn$_{12}$ is rather surprising since from the overlapping spin multiplets one
might have expected strong mixing effects. The explanation is found by realizing that transition matrix
elements are not only subject to the selection rules arising from the spin rotational symmetry (such as $|S
- S'| = 0, 1, 2$ for the spin mixing due to a bilinear anisotropy), but also to those arising from the
spatial, spin permutational properties of the microscopic spin Hamiltonian.\cite{OW_SYM,OW_INS} In fact, for
the Mn$_{12}$ cluster, or more precisely, for the microscopic spin Hamiltonian assumed for it,\cite{Cha04} it
turns out that the two lowest lying excited spin multiplets have a different spatial symmetry than the $S=10$
ground-state multiplet, and thus cannot mix with it [so that the quantities $\bar{D}_k$, which appear in both
the effective Hamiltonian, eq.~(\ref{HeffG}), and the effective spin operators, eq.~(\ref{Seff}), are zero
for these states]. Obviously, a high molecular symmetry can effectively switch off a mixing with many of the
higher lying spin multiplets. This teaches the important lesson that the spatial symmetry (or coupling
topology) of the cluster plays an important role in the spin mixing - and in all the effects related to it.
With respect to the magnetic anisotropy of the cluster, i.e., the effective spin Hamiltonian, this implies
that a deviation of the actual molecule from an assumed high symmetry can lead to substantially different
values of the parameters. This effect should be suspected as an important origin for higher-order terms in
the effective spin Hamiltonian (and for Mn$_{12}$ orthorhombic distortions in fact were demonstrated
recently, see Ref.~\onlinecite{Bir04} and references cited therein). Though significant spin mixing is
absent in the assumed model of Mn$_{12}$, it was discussed here as an example as it allows these important
and general conclusions concerning the role of spin mixing and the origin of magnetic anisotropy.

%

\section{Electron paramagnetic resonance}
\label{sEPR}

In the previous section, and in eq.~(\ref{HeffG}) of section~\ref{sHeff}, we have considered a system in zero
magnetic field under rather general conditions. In this section, we additionally take into account the Zeeman
term in the microscopic Hamiltonian, eqs.~(\ref{H}) or (\ref{Hito}). Otherwise, we make the same assumptions
as before, i.e., we assume $\hat{H}_1 = \sum_{ij} \hat{{\bf S}}_i \cdot {\bf \Lambda}_{ij} \cdot \hat{{\bf
S}}_j + \mu_B \sum_{i} \hat{{\bf S}}_i \cdot {\bf g}_{i} \cdot {\bf B}$ with traceless and diagonal ${\bf
\Lambda}_{ij}$ (we will refer to the two terms in $\hat{H}_1$ as $\hat{H}_D$ and $\hat{H}_Z$, respectively).

First, we discuss the effective spin Hamiltonian. Equation~(\ref{Heff2nd}) of section~\ref{sHeff} can be
applied directly. The only new point which needs to be addressed here is that $\hat{H}_1$ now consists of two
terms instead of one. Accordingly, the first-order part of the effective Hamiltonian $\hat{\bar{H}}$ also
consists of two terms, namely $\hat{{\bf S}} \cdot \bar{{\bf D}} \cdot \hat{{\bf S}}$ and $\mu_B \hat{{\bf
S}} \cdot \bar{{\bf g}} \cdot {\bf B}$ (these two terms will be denoted symbolically as $SDS$ and $SgB$),
where $\bar{{\bf D}} = \sum_{ij} \Gamma_2(ij) {\bf \Lambda}_{ij}$ and $\bar{{\bf g}} = \sum_i \Gamma_1(i)
{\bf g}_i$. For the second-order part, however, the four terms $P \hat{H}_D P' \hat{H}_D P$, $P \hat{H}_Z P'
\hat{H}_Z P$, $P \hat{H}_D P' \hat{H}_Z P$, and $P \hat{H}_Z P' \hat{H}_D P$ have to be evaluated.

The term $P \hat{H}_D P' \hat{H}_D P$ has been calculated already in section~\ref{sHeff}, eq.~(\ref{HeffG}).
It renormalizes the bilinear zero-field-splitting term (the symbol $SDS$ shall be understood as to include
this effect) and produces fourth-order terms (abbreviated symbolically as $S^4$).

The term $P \hat{H}_Z P' \hat{H}_Z P$, which is quadratic in the magnetic field, adds a $k=0$ and $k=2$
contribution [and a $k=1$ contribution, which, however, is related to vector products ${\bf b}_i \times {\bf
b}_j$ and thus smaller by a factor $(\delta g/g_0)^2$; it is neglected here]. The $k=0$ contribution produces
what is known as Van-Vleck or temperature-independent paramagnetism (TIP). It may be written as $-{1 \over 2}
\chi_0 B^2$ (or symbolically as $B^2$). The TIP parameter $\chi_0$ can be calculated straightforwardly from
eq.~(\ref{Heff2nd}). The $k=2$ contribution can be put in the form $\hat{{\bf S}} \cdot {\bf A} \cdot
\hat{{\bf S}} $, with the tensor ${\bf A}$ composed of the (reduced) magnetic fields only: ${\bf A} =
\sum_{ij} {\bf b}_i \cdot {\bf b}^T_j$. Thus, this contribution adds to the bilinear anisotropy, but the
anisotropy depends on the field direction and magnitude (it is denoted symbolically as $SB^2S$)

The terms $P \hat{H}_D P' \hat{H}_Z P$ and $P \hat{H}_Z P' \hat{H}_D P$, which both are linear in the
magnetic field, produce a $k=1$ and $k=3$ contribution (the $k=2$ term cancels out as wave functions may be
chosen as real). The $k=1$ contribution, since it is linear in both the magnetic field and the spin
operators, adds a term $\mu_B \hat{{\bf S}} \cdot \Delta{\bf g} \cdot {\bf B}$ to the effective Hamiltonian
[where $\Delta{\bf g}$ can be obtained again straightforwardly from eq.~(\ref{Heff2nd})]. Thus, this part
modifies the effective $g$ tensor, making it more anisotropic (the symbol $SgB$ shall be understood as to
include this effect). The $k=3$ contribution, however, is unusual as it is of third order in the spin
operators. It cannot be put into a bilinear form as the other terms (it is symbolically written as $S^3B$).

Altogether, the second-order effective spin Hamiltonian consists of the six terms $SDS$, $S^4$, $SB^2S$,
$S^3B$, $SgB$, and $B^2$. The above considerations seem pretty obvious, but appear to have been disregarded
in all actual analyses of EPR results on molecular spin clusters. The effective spin Hamiltonian used there
consisted of the zero-field-splitting terms (up to any desired order) and the Zeeman term (with a possibly
anisotropic g tensor). That is, in the present context only the terms $SDS$, $S^4$, and $SgB$ have been
considered. The $B^2$ term does not affect energy differences and thus can be omitted in spectroscopic work
(though it has been demonstrated to be of importance for a correct interpretation of high-field thermodynamic
data \cite{OW_Cu3x3}). However, we do not see a justification for omitting the remaining field-dependent
terms $SB^2S$ and $S^3B$, especially in high-field EPR (HFEPR).

In low-field EPR ($X$- and $Q$-band EPR), the terms $SB^2S$ and $S^3B$ might affect the analysis little, at
least one can hope so. But in HFEPR, which is actually motivated by the aim to work in the limit $\hat{H}_Z
\gg \hat{H}_D$, the effects introduced by these two terms should be significant. This might be objected to
with the general reasoning that these terms, as they are of second order in the perturbation theory, are
reduced by $1/\Delta$ and thus always small. However, whenever there are reasons to expect that e.g. the
parameters $B^m_4$ are significantly influenced by spin mixing (as it is often the case and has been
demonstrated explicitly for e.g. the single-molecule magnets Fe$_8$, Fe$_4$, and the molecular wheel Cr$_8$
\cite{Liv02,Car03,Car04}), one also should expect significant effects from the terms $SB^2S$ and $S^3B$.

A more detailed analysis of the resulting EPR spectra would be beyond the scope of this work even for the
simplest situations. We want to close with some comments.

From the above discussion it should be clear that with the current practice of neglecting non-linear field
dependencies in the spin-Hamiltonian, the parameter values determined by HFEPR and INS may differ
substantially (for the system Fe$_8$, e.g., it has been reported that the axial parameter $D$ as determined
by INS is 6\% larger than that obtained by HFEPR \cite{Cac98}). However, the non-Zeeman like field
dependence of the operators, in turn could be developed into a powerful tool to analyze the many-spin
effects: a careful comparison of spectra recorded at different fields/frequencies provides direct
information about the second-order field-dependent terms in the effective spin Hamiltonian, and thereby
about the effects due to spin mixing.

The calculation of the effective spin operators $\hat{\bar{S}}_{i\alpha}$ is simpler than the calculation of
$\hat{\bar{H}}$ in the sense, that the second-order contributions to $\hat{\bar{S}}_{i\alpha}$ can be
calculated independently for each term in $\hat{H}_1$, and then be added up. Thus, the results of the
previous section, eq.~(\ref{Seff}), provide directly the second-order contributions due to $\hat{H}_D$. The
calculation of the second-order contribution of the Zeeman term is, by following the same line as in the
previous section, easy enough, but the result is again pretty lengthy and therefore not reported here.

As for INS, a general result is possible for the EPR intensity if one considers only the first-order
contributions in the effective spin operators. In linear-response theory, the absorbed EPR power is given by
\begin{equation}
\label{epr}
  P({\bf B}) =
  C(\omega,T) \sum_{nm} {e^{-\beta E_n} \over Z(T)} P_{nm}({\bf B}) \delta(\omega-{E_m-E_n\over \hbar}),
\end{equation}
and
\begin{equation}
\label{eprnm}
  P_{nm}({\bf B}) = \sum_{ij} \sum_{\alpha \beta} b'_{i \alpha} b'_{j \beta}
  \langle n|\hat{S}_{i \alpha}|m\rangle\langle m| \hat{S}_{j \beta}|n\rangle,
\end{equation}
with $C(\omega,T) = \omega  ( 1- e^{- \beta \hbar \omega} ) / (8\pi\hbar)$, and the (reduced) microwave field
${\bf b}'_i = \mu_B {\bf g}_i \cdot {\bf B}'$ (which is to be contrasted to the static magnetic field ${\bf
B}$, or ${\bf b}_i$). Insertion of $\hat{\bar{S}}_{i \alpha} = \Gamma_1(i) \hat{S}_{\alpha}$ leads directly
to the first-order result
\begin{equation}
\label{epr1st}
  P^{1st}_{nm}({\bf B}) = \sum_{\alpha \beta} b'_{\alpha} b'_{\beta}
  \langle n|\hat{S}_{\alpha}|m\rangle\langle m| \hat{S}_{\beta}|n\rangle,
\end{equation}
with the field ${\bf b}' = \mu_B {\bf \bar{g}} \cdot {\bf B}'$ [we note that for the case ${\bf g}_i = g
\mathbf{1}$, the EPR intensity becomes proportional to the INS intensity at $Q = 0$, see eq.~(\ref{I1st})].
Thus, as expected and mentioned at the end of section~\ref{sSSApproach}, the many-spin nature of the wave
functions does not affect the relative intensities of the EPR transitions in first-order perturbation theory.
Up to this order, the single-spin approach is justified. But as in the case of INS, second-order effects due
to spin mixing can be significant too. However, in view of the inaccuracies introduced already by the use of
an incomplete spin Hamiltonian in the analysis of (HF)EPR results, see the above discussion, we are not
pursuing this issue further here.

%

\section{Conclusions}
\label{sConclusions}

In this work, we have investigated the influence of spin mixing, i.e., a mixing between the different spin
multiplets in the spectrum of a Heisenberg spin system, on the spectroscopic intensities of transitions
within a particular spin multiplet. This situation is encountered in many molecular nanomagnets, with
systems such as the single-molecule magnets as canonical examples. The spectroscopic intensities were
calculated perturbationally, with the isotropic Heisenberg exchange terms as the unperturbed part. Spin
mixing arises then from the anisotropic terms in the microscopic spin Hamiltonian, and its effects were
calculated up to second-order in perturbation theory. The consequent use of the spherical tensor and ITO
formalism yielded a number of useful, and easily evaluated, equations. The resulting mixing effects were
discussed for the specific examples of INS and EPR, with strong emphasis on the former technique.

As a main result it was observed that the transition intensities are affected individually for each
transition, as compared to the results of a first-order calculation or the single-spin approach,
respectively, which has been the approach used in all spectroscopic studies on the splitting in spin ground
states so far. The transition intensities in fact were found to provide additional information, which in
some cases may not be obtained otherwise. For instance - and this is probably the most important example for
that - we showed that the careful inclusion of the transition intensities in the analysis of the observed
spectra allows to differentiate between the two origins of higher-order magnetic anisotropy terms in the
ground states, namely spin mixing or single-ion higher-order terms. Such studies would thus allow an
unprecedented insight into the microscopic origin of magnetic anisotropy in spin clusters, which ultimately
could help to achieve a tailored design of magnetic molecules. We also showed, that the symmetry of the
cluster, or its absence, can have a profound impact on the resulting anisotropy of the spin clusters.

As a summary, in this work we established a complete toolbox for the analytical analysis of the spin-mixing
effects, including a general equation for the construction of the second-order effective spin Hamiltonian,
and explored the various resulting effects, as well as the kind of information which may be obtained from
them.

Finally, we should address the experimental feasibility of such investigations. The considered examples
showed that changes in the intensities due to spin mixing may be on the order of 10\% or so. With the INS
spectrometers available nowadays, such effects should be detectable for magnetically strongly scattering
species, which would be typically low-nuclearity spin clusters like dimers. Such systems in fact would be
excellent model compounds for a demonstration of the proposed mixing effects or development of the
methodology. For the most interesting molecules, however, like the single-molecule magnets, such studies seem
to be prohibitive, if not impossible, at the moment. For EPR the situation is rather similar. However,
considering the enormous improvements of the INS and EPR techniques in the last decade, the outlined studies
should become possible in the future. We hope, that this work will stimulate development efforts in this
direction - the new opportunities which would be gained, in our opinion, would more than reward these
efforts.

%

\appendix

\section{}
\label{app_ITO}

Some comments shall be added concerning the Cartesian and spherical representations of a spin Hamiltonian.
The general form of each term is predetermined by the fact that a Hamiltonian has to transform as a scalar
under rotations in space. The scalar product of two spherical tensors has the general form $\sum_{q}
U^{(k)*}_{q} V^{(k)}_q$, or $\sum_{q} (-1)^q U^{(k)}_{-q} V^{(k)}_q$ since $U^{(k)*}_{q} = (-1)^q
U^{(k)}_{-q}$ (here $U$ and $V$ are arbitrary spherical tensors, e.g., ITOs). The reader may notice this
motive in many equations in this manuscript. In the ITO representation,\cite{ITO} the transformation
properties thus enforce the spin Hamiltonian to appear - in the notation of this work - as a sum of terms
like $\sum_{q}(-1)^q T^{(k)}_{-q} \hat{T}^{(k)}_q(s)$, i.e., as a sum of spherical scalar products.

For instance, for a bilinear term follows
\begin{equation}
\label{app1}
 \hat{{\bf S}}_i \cdot {\bf \Lambda} \cdot \hat{{\bf S}}_j = \sum_{k=0,1,2} \sum_{q} T^{(k)*}_{q}({\bf
\Lambda}) \hat{T}^{(k)}_q(ij),
\end{equation}
The association of the $k=0, 1, 2$ contributions in this expression with the isotropic, anti-symmetric, and
symmetric\&traceless parts of the (Cartesian) tensor ${\bf \Lambda}$ is standard. The generalization of this
concept to the other terms in the spin Hamiltonian is obvious. For instance, the Zeeman term $\mu_B
\hat{{\bf S}}_i \cdot {\bf g}_i \cdot {\bf B}$, which can be written as $\hat{{\bf S}}_i \cdot {\bf b}_i$
with ${\bf b}_i = \mu_B {\bf g}_i \cdot {\bf B}$, has to become the scalar product of the spherical tensor
$T^{(1)}_{q}({\bf b}_i)$ and the tensor operator $\hat{T}^{(1)}_{q}(i)$, both of rank one.

Any arbitrary operator $\hat{O}(S)$, which is a function of spin operators, can be expanded into ITOs
according to $\hat{O}(S) = \sum_{kq} c_{kq} \hat{T}^{(k)}_q(S)$. Section~\ref{sGeneralB} provides a general
procedure to find the coefficients $c_{kq}$. For the bilinear term, eq.~(\ref{app1}), one would find $c_{kq}
= (-1)^q T^{(k)}_{-q}({\bf \Lambda})$ in full agreement with the result concluded from the transformation
properties.

The bilinear term, eq.~(\ref{app1}), shall be discussed further. Let us consider a diagonal and traceless
tensor ${\bf \Lambda}$ [that is, the $k=0,1$ contributions in eq.~(\ref{app1}) are zero]. The components of
the spherical tensor $T^{(2)}_{q}({\bf \Lambda})$ associated with ${\bf \Lambda}$ are $T^{(2)}_0({\bf
\Lambda}) = (2\Lambda_{zz}-\Lambda_{xx}-\Lambda_{yy})/\sqrt{6}$ and $T^{(2)}_2({\bf \Lambda}) =
T^{(2)}_{-2}({\bf \Lambda}) = (\Lambda_{xx}-\Lambda_{yy})/2$. The $q = \pm 1$ components vanish. This
suggests to define the parameters
\begin{eqnarray}
 D &=& (2\Lambda_{zz}-\Lambda_{xx}-\Lambda_{yy})/2,
 \cr
 E &=& (\Lambda_{xx}-\Lambda_{yy})/2,
\end{eqnarray}
so that $T^{(2)}_0({\bf \Lambda}) = \sqrt{2/3} D$ and $T^{(2)}_{\pm 2}({\bf \Lambda}) = E$, in order to
parameterize the interaction described by $\hat{{\bf S}}_i \cdot {\bf \Lambda} \cdot \hat{{\bf S}}_j$.

In Refs.~\onlinecite{Bor99} and \onlinecite{Car03}, ${\bf \Lambda}$ was parameterized by $J^v =
(2\Lambda_{zz}-\Lambda_{xx}-\Lambda_{yy})/\sqrt{6}$ and $J^u = (\Lambda_{xx}-\Lambda_{yy})/\sqrt{2}$. $J^v$
agrees with $T^{(2)}_0({\bf \Lambda})$, but $J^u$ differs from $T^{(2)}_{\pm 2}({\bf \Lambda})$ by a factor
of $\sqrt{2}$. We prefer our choice of parameters because of two reasons. Firstly, it is consistent with the
usually chosen parametrization $\Lambda_{xx} = - 1/3 D + E$, $\Lambda_{yy} = - 1/3 D - E$, and $\Lambda_{zz}
= 2/3 D$ \{which for $i=j$ leads to the spin operators $D [\hat{S}^2_{iz}- {1 \over 3}S_i(S_i+1)]$ and
$E(\hat{S}^2_{ix}-\hat{S}^2_{iy})$\}. Secondly, it ensures the above compact form eq.~(\ref{app1}) of the
Hamiltonian, which simplifies calculations enormously as the rotational invariance properties are inherently
manifest.

\begin{table}
\caption{\label{tab2}Some irreducible tensor operators which are often needed in practice. $\hat{O}^0_4$,
$\hat{O}^2_4$, and $\hat{O}^4_4$ are Stevens operators.\cite{Abr70}}
\begin{ruledtabular}
\begin{tabular}{lll}
  $\hat{T}^{(2)}_0$  &=& $\sqrt{3 \over 2} \: [ \hat{S}^2_z - 1/3S(S+1)]$  \cr
  $\hat{T}^{(2)}_2 + T^{(2)}_{-2}$ &=& $\hat{S}^2_x-\hat{S}^2_y$ \\
\hline
  $\hat{T}^{(3)}_0$ &=& ${1 \over \sqrt{10}} \: [5 \hat{S}_z^2 - 3S(S+1) +1]\hat{S}_z$ \cr
  $\hat{T}^{(3)}_{\pm1}$ &=& $\mp \sqrt{3 \over 40} \: [5 \hat{S}_z^2 - 5 \hat{S}_z - (S-1)(S+2)]\hat{S}_{\pm}$ \\
  $\hat{T}^{(3)}_2 + T^{(3)}_{-2}$ &=&  $\sqrt{3} \: (\hat{S}_z-1)( \hat{S}_x^2 - \hat{S}_y^2 )$ \cr
  $\hat{T}^{(3)}_{\pm3}$ &=& $\mp {1 \over 2 \sqrt{2}} \: \hat{S}^3_{\pm}$ \\
\hline
  $\hat{T}^{(4)}_0$ &=& ${1 \over 2 \sqrt{70}} \: \hat{O}^0_4$   \cr
  $\hat{T}^{(4)}_2 + T^{(4)}_{-2}$ &=& ${1 \over \sqrt{7}} \: \hat{O}^2_4$  \cr
  $\hat{T}^{(4)}_4 + T^{(4)}_{-4}$ &=& ${1 \over 2} \: \hat{O}^4_4$
\end{tabular}
\end{ruledtabular}
\end{table}

\section{}
\label{app_TAB}

Table~\ref{tab2} lists the most often needed ITOs explicitly expressed in operator form. With these
expressions and the Wigner 3$j$ symbol
\begin{eqnarray}
  \left(\begin{array}{ccc} S&S&k \\ S&-S&0 \end{array} \right) = \sqrt{ (2S)! / (2S-k)! \over (2S+1+k)! /
  (2S)! }
\end{eqnarray}
it is straight forward to calculate the reduced matrix elements $\langle S \| \hat{T}^{(3)}(S) \| S \rangle$
and $\langle S \| \hat{T}^{(4)}(S) \| S \rangle$ not usually found in textbooks. For $k \leq 4$ one obtains
$\langle S \| \hat{T}^{(k)}(S) \| S \rangle = \sqrt{ (2S+1+k)! \over (2S-k)! \, t_k }$ with $t_0 = 1$, $t_1 =
4$, $t_2 = 24$, $t_3 = 160$, and $t_4 = 1120$.

\section{}
\label{app_INS}

For calculations an explicit expression of the INS formula eq.~(\ref{Inm}) is often needed, which is given
here for convenience. With $I_{nm}(Q) = \sum_{ij} F^*_i(Q) F_j(Q)  I_{nm,ij}(Q)$, eq.~(\ref{Inm}) reads
\begin{widetext}
\begin{eqnarray}
\label{Inmexp}
 I_{nm,ij}(Q) &=&
 {2 \over 3} j_0(Q R_{ij})
    ( \tilde{S}_{ix} \tilde{S}_{jx} + \tilde{S}_{iy} \tilde{S}_{jy} + \tilde{S}_{iz} \tilde{S}_{jz} )
  + {1 \over 6} j_2(Q R_{ij}){3 R^2_{ij,z} - R^2_{ij} \over R^2_{ij}}
       (2 \tilde{S}_{iz} \tilde{S}_{jz} - \tilde{S}_{ix} \tilde{S}_{jx} - \tilde{S}_{iy} \tilde{S}_{jy} )
  \cr &&
  + {1 \over 2} j_2(Q R_{ij}){R^2_{ij,x} - R^2_{ij,y} \over R^2_{ij}}
      ( \tilde{S}_{ix} \tilde{S}_{jx} - \tilde{S}_{iy} \tilde{S}_{jy}  )
 + j_2(Q R_{ij}){R_{ij,x} R_{ij,y} \over R^2_{ij}}
       ( \tilde{S}_{ix} \tilde{S}_{jy} + \tilde{S}_{iy} \tilde{S}_{jx}  )
  \cr &&
  + j_2(Q R_{ij}){R_{ij,x} R_{ij,z} \over R^2_{ij}}
       ( \tilde{S}_{ix} \tilde{S}_{jz} + \tilde{S}_{iz} \tilde{S}_{jx}  )
 + j_2(Q R_{ij}){R_{ij,y} R_{ij,z} \over R^2_{ij}}
       ( \tilde{S}_{iy} \tilde{S}_{jz} + \tilde{S}_{iz} \tilde{S}_{jy}  ).
\end{eqnarray}
\end{widetext}
For real wave functions, which is the situation mostly encountered in praxis, the fourth and sixth term on
the r.h.s. vanish and can be dropped.

It is actually often useful to have this formula in another form. Since the ${\bf R}_{ij}$-dependent
prefactors do not change upon the substitution $ij \rightarrow ji$, one may replace each $\tilde{S}_{i\alpha}
\tilde{S}_{j\beta}$ by $(\tilde{S}_{i\alpha} \tilde{S}_{j\beta} + \tilde{S}_{j\alpha}
\tilde{S}_{i\beta})/2$. If one further expresses the spin operators $\hat{S}_{ix}$ and $\hat{S}_{iy}$ by the
ladder operators $\hat{S}_{i}^{\pm}$, uses $\langle n| \hat{S}_{i}^{\pm} | m \rangle = \langle m|
\hat{S}_{i}^{\mp} | n \rangle^*$, and introduces the abbreviations ${i}_{+} \equiv \langle n|
\hat{S}_{i}^{+} | m \rangle$ and similar for the $z$ and $-$ components, one obtains the handy equation
\begin{widetext}
\begin{eqnarray}
\label{Inmexp2}
 I_{nm,ij}(Q) &=&
 {1 \over 3} j_0(Q R_{ij})
        \mathrm{Re}[ 2 i_z j_z^* + i_+ j_+^* +  i_- j_-^* ]
  + {1 \over 12} j_2(Q R_{ij}){3 R^2_{ij,z} - R^2_{ij} \over R^2_{ij}}
        \mathrm{Re}[4 i_z j_z^* - i_+ j_+^* - i_- j_-^* ]
  \cr &&
  + {1 \over 2} j_2(Q R_{ij}){R^2_{ij,x} - R^2_{ij,y} \over R^2_{ij}}
      \mathrm{Re}[ i_+ j_-^* ]
 + j_2(Q R_{ij}){R_{ij,x} R_{ij,y} \over  R^2_{ij}}
      \mathrm{Im}[ i_+ j_-^* ]
  \cr &&
  + j_2(Q R_{ij}){R_{ij,x} R_{ij,z} \over  R^2_{ij}}
     \mathrm{Re}[ (i_+ + i_-) j_z^* ]
 + j_2(Q R_{ij}){R_{ij,y} R_{ij,z} \over R^2_{ij}}
     \mathrm{Im}[ (i_+ - i_-) j_z^* ],
\end{eqnarray}
\end{widetext}
where $\mathrm{Re}[]$ and $\mathrm{Im}[]$ denote the real and imaginary parts, respectively. This form of
eq.~(\ref{Inm}) is also useful if the spin matrix elements are calculated by the ITO technique\cite{Bor99}
since the corresponding ITO expressions $i_z = \langle n| \hat{T}^{(1)}_0(i) | m\rangle$ and $i_{\pm} = \mp
\sqrt{2} \langle n| \hat{T}^{(1)}_{\pm} (i) | m\rangle$ are readily inserted.


%

\begin{acknowledgments}
We thank R. Bircher, S. T. Ochsenbein, and A. Sieber for many useful comments and suggestions. OW gratefully
acknowledges financial support by EC-RTN-QUEMOLNA, contract n$^\circ$ MRTN-CT-2003-504880.
\end{acknowledgments}

%

%

\begin{references}
\bibitem{FDMRS}
In recent years, also the zero-field version of EPR, called frequency-domain magnetic resonance
spectroscopy, has been applied successfully.\cite{Sla03} Being a zero-field technique, it shares analogies
with INS, but will not be discussed explicitly here since everything derives in an obvious manner from what
is found for INS.
\bibitem{TOF}
The experimental intensity of a peak at energy $E_{mn}$ as measured in channel $k$, $I_k(E_{mn})$,
corresponds to $I_{mn}(Q=Q_k)$, where $Q_k$ is the momentum transfer in channel $k$ at a given energy
transfer $E_{mn}$. For the integrated TOF intensity $I_{nm}^{tot} \equiv \sum_k I_k(E_{mn})$ thus holds:
$I_{nm}^{tot} = \sum_k I_{mn}( Q_k ) = \sum_{\Delta Q} I_{mn}( Q ) (\Delta k / \Delta Q) \Delta Q =
\int_{Q_{min}}^{Q_{max}} g(Q) I_{nm}(Q) dQ$. Here, $Q_{min}$ and $Q_{max}$ depend via the scattering triangle
on the energy transfer and the minimal and maximal angles covered by the detector bank, and $g(Q)$ is the
density of detector banks per $Q$ range.
\bibitem{EPR1}
If the $g$ factors of the individual spin centers in a hetero-nuclear cluster differ significantly, then in
fortunate cases EPR can provide some information on the spatial distribution of the transition strength. A
recent example is provided in Ref.~\onlinecite{Cad04}.
\bibitem{ITO}
In this work, the spherical tensors and ITOs were defined as in Ref.~\onlinecite{Mes85}, consistent with most
literature sources. All definitions found in the literature fulfill the equations $[\hat{S}_z,T^{(k)}_q] = q
T^{(k)}_q$ and $[\hat{S}_{\pm},T^{(k)}_q] = \sqrt{k(k+1)-q(q \pm 1)} T^{(k)}_{q \pm 1}$, which ensure the
correct transformation properties. Unfortunately, these equations define the $T^{(k)}_q$ only up to a
constant factor (which might be the reason for the confusing number of definitions of spin operator
equivalents). For the results in this work to hold, the spherical tensors and ITOs have to fulfill also the
tensorial product equation $T^{(K)}_{Q} \equiv [U^{(k1)} \otimes V^{(k2)}]^{(K)}_Q = \sum_{q1,q2} \langle k_1
k_2 q_1 q_2 | K Q \rangle U^{(k1)}_{q_1} V^{(k2)}_{q_2}$ (this equation is not obeyed by, e.g., the spherical
harmonics $Y_{lm}$).
\bibitem{ITO2}
The composed ITOs are not completely specified by our notation $\hat{T}^{(k)}_q(ij)$. For an ITO composed of
two single-spin ITOs $\hat{U}^{(k_u)}_{q_u}(i)$ and $\hat{V}^{(k_v)}_{q_v}(j)$, which act in the single-spin
spaces of the $i$th and $j$th center, a precise notation would require a bookkeeping of the individual ranks
$k_u$ and $k_v$, e.g., by writing $\hat{T}^{(k)}_q(k_uk_v|ij)$ (this notation has been used in
Refs.~\onlinecite{Bor99} and \onlinecite{Car04}). $k_u$ and $k_v$ of course are involved in actual
calculations, but are always clear from the context of the considered term, and are thus suppressed here in
order to avoid unnecessary clutter in the equations.
\bibitem{sparse}
In these numerical calculations, the relevant eigenpairs were obtained by the subspace iteration method.
Whereby, $S_z$ symmetry, where $S_z$ is the $z$ component of the total spin operator, was employed as usual.
\end{references}
\end{document}